\documentclass[3p,12pt]{elsarticle}

\usepackage{amssymb}
\usepackage{amsmath,amsfonts}
\usepackage{url}
\usepackage{graphicx}
\usepackage{framed,multirow}
\usepackage{lineno}
\usepackage{color}

\usepackage{subfig}
\newcommand{\bx}{\boldsymbol{x}}
\newcommand{\bp}{\boldsymbol{p}}
\newcommand{\acc}{\textsf{acc}}
\newcommand{\conf}{\textsf{conf}}
\def\bbeta{{\boldsymbol{\beta}}}
\newcommand{\calR}{\mathcal{R}}
\newcommand{\loss}{\textsf{loss}}
\newcommand{\BS}{\textsf{BS}}
\newcommand{\NLL}{\textsf{NLL}}

\newcommand{\ECE}{\textsf{ECE}}
\newcommand{\new}[1]{{\color{black}{#1}}} 
\newcommand{\revision}[1]{{\color{black}{#1}}} 
\journal{Information Fusion(Accepted)}
\newtheorem{Rem}{Remark}
\newtheorem{Ex}{Example}

\begin{document}

\begin{frontmatter}

\title{Deep evidential fusion with uncertainty quantification and reliability learning for multimodal medical image segmentation}

\author[inst1]{Ling Huang}
%\ead{iweisskohl@gmail.com}
%\cortext[cor1]{Corresponding author: Ling Huang, Email: iweisskohl@gmail.com}
\affiliation[inst1]{organization={Université de technologie de Compiègne, CNRS, Heudiasyc},%Department and Organization
%            addressline={Rue Roger Couttolenc}, 
            city={Compiègne},
%            postcode={60200}, 
            %city={CompiÃ¨gne},
            country={France}}

\author[inst2]{Su Ruan}

\affiliation[inst2]{organization={Université de Rouen Normandie, Quantif, LITIS},%Department and Organization
%            addressline={Rue d'Amiens}, 
            %city={City Two},
%            postcode={76038}, 
            city={Rouen},
            country={France}}
\author[inst3]{Pierre Decazes}
\affiliation[inst3]{organization={Université de Rouen Normandie, Centre Henri Becquerel},%Department and Organization
%            addressline={Rue dâAmiens}, 
            %city={City Two},
%            postcode={76038}, 
            city={Rouen},
            country={France}}
\author[inst1,inst4]{Thierry Den{\oe}ux}
\affiliation[inst4]{organization={Institut universitaire de France},%Department and Organization
%            addressline={103 Saint-Michel}, 
            %city={City Two},
%            postcode={75005}, 
            city={Paris},
            country={France}}
\begin{abstract}
Single-modality medical images generally do not contain enough information to reach an accurate and reliable diagnosis. For this reason, physicians commonly rely on multimodal medical images for comprehensive diagnostic assessments. This study introduces a deep evidential fusion framework designed for segmenting multimodal medical images, leveraging the Dempster-Shafer theory of evidence in conjunction with deep neural networks. In this framework, features are first extracted from each imaging modality using a deep neural network, and features are mapped to Dempster-Shafer mass functions that describe the evidence of each modality at each voxel. The mass functions are then corrected by the contextual discounting operation, using learned coefficients quantifying the reliability of each source of information relative to each class.  The discounted evidence from each modality is then combined using Dempster's rule of combination. Experiments were carried out on a PET-CT dataset for lymphoma segmentation and a multi-MRI dataset for brain tumor segmentation. The results demonstrate the ability of the proposed fusion scheme to quantify segmentation uncertainty and improve segmentation accuracy. Moreover, the learned reliability coefficients provide some insight into the contribution of each modality to the segmentation process.
\end{abstract}

\iffalse
\begin{graphicalabstract}
\includegraphics[width=\textwidth]{figure/global_arch.pdf}
\end{graphicalabstract}

\begin{highlights}
\item A new deep framework for multi-modality medical image segmentation is introduced.
\item The approach is based on Dempster-Shafer theory of evidence.
\item Mass functions are transformed by contextual discounting with learned coefficients.
\item The discounting coefficients provide information about the fusion process.
\item The model performance is assessed using two PET-CT and multi-MRI datasets.
\end{highlights}
\fi


\end{frontmatter}

%\linenumbers

\section{Introduction}
\label{sec: intro}

Recent advances in medical imaging technologies have facilitated the acquisition of multimodal data such as Positron Emission Tomography (PET)/Computed Tomography (CT) and multi-sequence Magnetic Resonance Imaging (MRI). Images from a single modality provide partial insight into cancer and other abnormalities within the human body. Multimodal medical image analysis, which integrates information from diverse medical imaging modalities, significantly contributes to a comprehensive understanding of intricate medical conditions \cite{zhang2020advances}. It encompasses factors such as the location, size, and extent of pathological structures. Medical image segmentation based on the fusion of multimodal medical information allows clinicians to better delineate anatomical structures, lesions and abnormalities, thus enhancing the effectiveness of disease detection, diagnosis, and treatment planning. 

Multimodal medical image fusion strategies can be implemented at different levels \cite{weng2024semi}. At the lowest pixel level,  multimodality images are concatenated as a single input.  Alternatively,  features can be extracted from different modalities and combined for further modeling and reasoning (feature-level fusion). Finally, in the decision-level approach,  partial decisions are made independently based on each modality and aggregated to obtain a final decision. Though recent developments in multimodal medical image analysis have yielded promising experimental results,  conventional multimodal medical image fusion strategies still suffer from some limitations. It is often difficult to explain why a given strategy works in a given context, and to quantify decision uncertainty in a reliable way. Moreover, most approaches are based on optimistic assumptions about data quality and, contrary to clinical knowledge, they treat images from different modalities as equally reliable when segmenting tumors,  which may lead to biased or wrong decisions.

The success of information fusion depends on the relevance and complementarity of input information, the existence of prior knowledge about the information sources, and the expressive power of the uncertainty model employed \cite{rogova2004reliability, dubois2016basic,pichon19}. Given that the quality of input information and prior knowledge is intricately tied to the data collection stage, a lot of work has been devoted to modeling uncertainties in a faithful way  \cite{huang2023review}. As a critical factor in the information fusion process \cite{abdar2021review, huang2022application}, accurate uncertainty quantification must be regarded as a primary objective to achieve precise multimodal medical image segmentation.

Early methods for quantifying uncertainty essentially relied on probabilistic models, often integrated with Bayesian inference or sampling techniques to estimate uncertainty across various parameters or variables \cite{hinton1993keeping, mackay1992practical}. The advent of deep neural networks has sparked renewed interest in uncertainty estimation \cite{abdar2021review}, leading to the development of methods such as Monte-Carlo dropout \cite{gal2016dropout} and deep ensembles \cite{lakshminarayanan2017simple}. However, it is important to note that these probabilistic models rely on assumptions about the underlying data distribution, and improper distributions can result in inaccurate uncertainty estimations. %Additionally, uncertainty quantification through inference or sampling algorithms relies heavily on statistical analysis and lacks theoretical explanation, posing limitations for information fusion and decision-making studies.
Furthermore, uncertainty quantification via inference or sampling algorithms heavily relies on computational approximations and may lack rigorous theoretical justification \cite{bauer2016understanding, bengio2017deep}. These and other limitations motivate the search for alternative approaches for uncertainty quantification for information fusion and decision-making applications.

Instead of making strong assumptions on actual data distribution, non-probabilistic methods use alternative mathematical frameworks or representations such as possibility theory \citep{zadeh78,denoeux20a} and Dempster-Shafer theory (DST) \cite{dempster1967upper, shafer1976mathematical, denoeux20b} to quantify uncertainty. In particular, the latter formalism is an evidence-based information modeling, reasoning, and fusion framework that can be used with both supervised \cite{denoeux2000neural,tong21a,tong21b} and unsupervised learning \cite{lian19,denoeux21b}, providing an effective way to handle imperfect (i.e., imprecise, uncertain, and conflicting) data. Compared to possibility theory, DST allows the quantification of both aleatory and epistemic uncertainty while providing a powerful mechanism for combining multiple unreliable pieces of information \cite{pichon19}.

In multimodal medical image segmentation, effectively combining uncertain information from diverse sources presents a significant challenge. Some learning-based approaches propose addressing conflicting decisions by introducing learnable weights \cite{liu2017medical, asha2019multi, shi2023uncertainty}. The term ``weight'' in those approaches usually refers to the importance of information. In contrast, reliability pertains to the trustworthiness of the information and needs to be carefully analyzed in different medical situations. Four major approaches have been used to provide reliability coefficients: 1) modeling the reliability of sources using a degree of consensus \cite{delmotte1996context}; 2) modeling expert opinions using probability distributions \cite{cooke1991experts}; 3) using external domain knowledge or contextual information to model reliability coefficients \cite{fabre2001presentation}; 4) learning the reliability coefficients from training data \cite{elouedi2004assessing,pichon16}, which is a very general approach that does not require any prior domain knowledge or expert opinions. In this work, we consider an even more flexible approach in which the reliability of each image modality is described by several coefficients, one for each ground truth value. The reliability coefficient for source $i$ and class $k$ is then defined as one's belief that the information from source $i$ is reliable, if the true class is $k$. 

In this paper, we introduce a new approach to multimodal medical image segmentation combining  DST with deep neural networks\footnote{This paper is an extended version of the short paper presented at the 25th International Conference on Medical Image Computing and Computer Assisted Intervention (MICCAI 2022) \cite{huang2022evidence}. This extended version includes a much more detailed description and explanation of the fusion framework, an improved optimization strategy with a two-part loss function, as well as extended results with a second dataset for lymphoma segmentation and an additional transformer-based feature-extraction module.}. The proposed fusion scheme comprises multiple encoder-decoder-based feature extraction modules, DST-based evidence-mapping modules, and a multimodality evidence fusion module. The evidence-mapping modules transform the extracted features into mass functions representing the evidence from each imaging modality about the class of each voxel. These mass functions are then corrected by a contextual discounting operation, and the discounted pieces of evidence are combined by Dempster's rule of combination. The whole framework is trained end-to-end by minimizing a loss function quantifying the errors before and after the fusion of information from each modality. Our main contributions are, thus, the following:
\begin{enumerate}
 \item We propose a new hybrid fusion architecture for multimodal medical images composed of feature extraction, evidence-mapping, and combination modules.
 \item Within this architecture, we integrate mechanisms for (i) quantifying segmentation uncertainty using Dempster-Shafer mass functions, (ii)  correcting these mass functions to account for the relative reliability of each imaging modality using context discounting, and (iii) combining corrected mass functions from different sources to reach final segmentation decisions.
 \new\item We  introduce an improved two-part loss function making it possible to optimize the segmentation performance of each individual source modality together with the overall performance of the combined decisions.
 \item Through extensive experiments with two real medical image datasets, we show that the proposed decision-level fusion scheme improves segmentation reliability and quality as compared to alternative pixel-level methods for exploiting different image modalities.
 \item We show that the learned reliability coefficients provide some insight into the contribution of each imaging modality in the segmentation process.
\end{enumerate}

The rest of this paper is organized as follows. Background information and related work are first recalled in Section \ref{sec: related}. Our approach is then introduced in Section \ref{sec: proposed}, and experimental results are reported in Section \ref{sec: exp}. Finally, Section \ref{sec:conclu} concludes the paper and presents some directions for further research.

\section{Related work}
\label{sec: related}

The basic concepts of DST and its application to classification are first  recalled in Section \ref{sec:dst}. The contextual discounting operation, which plays a central role in our approach, is described separately in Section \ref{subsec: discount}. The evidential neural network model used in this paper is then introduced in Section \ref{subsec: enn}, and related work on multimodal medical image fusion is briefly reviewed in Section \ref{subsec:multimod}.

\subsection{Dempster-Shafer theory}
\label{sec:dst}
Let $\Theta =\{\theta _{1},\theta _{2}, \ldots, \theta_{K}\} $ be the finite set of possible answers to some question, called the \emph{frame of discernment}. Evidence about a variable taking values in $\Theta$ can be represented by a \emph{mass function} $m: 2^{\Theta}\rightarrow [0, 1]$, such that
%\begin{subequations}
\begin{equation*}
%\label{eq:mass}
    \sum _{A\subseteq \Theta }m(A)=1 \quad \text{and} \quad 
     m(\emptyset)=0.
\end{equation*}
%\end{subequations}
Each subset $A \subseteq \Theta$ such that $m(A)>0$ is called a \emph{focal set} of $m$. The mass $m(A)$ represents a share of a unit mass of belief allocated to focal set $A$, which cannot be allocated to any strict subset of $A$. The mass $m(\Theta)$ can be interpreted as a degree of ignorance. Full ignorance is represented by the \emph{vacuous} mass function $m_?$ verifying $m_? (\Theta)=1$. If all focal sets are singletons, then $m$ is said to be \emph{Bayesian}; it is equivalent to a probability distribution.

\paragraph{Belief and plausibility functions} The information provided by a mass function $m$ can also be represented by a \emph{belief function} $Bel$ or a \emph{plausibility function} $Pl$ from $2^{\Theta }$ to $[0,1]$ defined, respectively, as: 
\begin{equation*}
   Bel(A) = \sum _{ B\subseteq A}m(B)
 %  \label{eq:bel}
\end{equation*}
and 
\begin{equation*}
   Pl(A) = \sum _{B\cap A\neq \emptyset }m(B)=1-Bel(\bar{A}),
%   \label{eq:plau}
\end{equation*}
for all $A\subseteq \Theta$, where $\bar{A}$ denotes the complement of $A$. The quantity $Bel(A)$ can be interpreted as a degree of support for $A$, while $Pl(A)$ is a measure of lack of support against $A$. The \emph{contour function} $pl$ associated to $m$ is the function that maps each element $\theta$ of $\Theta$ to its plausibility, i.e.,
\begin{equation*}
    pl(\theta)=Pl(\{ \theta\}),  \quad \forall \theta \in \Theta.
\end{equation*}
As shown below, this function can be easily computed when combining several pieces of evidence; it plays an important role in decision-making.

\paragraph{Dempster's rule}
\label{subsubsec: dempster}
In DST, the beliefs about a certain question are established by aggregating independent pieces of evidence represented by belief functions over the same frame of discernment \cite{shafer1976mathematical}. Given two mass functions $m_{1}$ and $m_{2}$ derived from two independent items of evidence, the mass function $m_{1}\oplus m_{2}$ representing the pooled evidence is defined as 
\begin{subequations}
\label{eq:demp1}
    \begin{equation}
    (m_{1}\oplus m_{2})(A)=\frac{1}{1-\kappa }\sum _{B\cap C=A}m_{1}(B)m_{2}(C),
\end{equation}
for all $A\subseteq \Theta, A\neq \emptyset$, and $(m_{1}\oplus m_{2})(\emptyset)=0$. The coefficient $\kappa$ is the \emph{degree of conflict} between $m_{1}$ and $m_{2}$, 
\begin{equation}
    \kappa=\sum _{B\cap C=\emptyset}m_{1}(B)m_{2}(C).
\end{equation}
\end{subequations}

This operation is called \emph{Dempster's rule of combination}. It is commutative and associative. The combined mass function $m_1 \oplus m_2$ is called the \emph{orthogonal sum} of $m_1$ and $m_2$. Mass functions $m_1$ and $m_2$ can be combined if and only if $\kappa<1$. Let $pl_1$, $pl_2$ and $pl_{1}\oplus pl_{2}$ denote the contour functions associated with, respectively, $m_1$, $m_2$ and $m_{1}\oplus m_{2}$. The following equation holds:
\begin{equation}
        \forall \theta\in\Theta, \quad (pl_{1}\oplus pl_{2}) (\theta)=\frac{pl_1 (\theta) pl_2(\theta)}{1-\kappa}.
        \label{eq:prodpl}
\end{equation}
The complexity of calculating the combined contour function using \eqref{eq:prodpl} is linear in the cardinality of $\Theta$, whereas computing the combined mass function using \eqref{eq:demp1} has, in the worst-case, exponential complexity.

\paragraph{Conditioning} Given a mass function $m$ and a nonempty subset $A$ of $\Theta$ such that $Pl(A)>0$, the conditional mass function $m(\cdot\vert A)$ is defined as the orthogonal sum of $m$ and the  mass function $m_A$ such that $m(A)=1$. Conversely, given a conditional mass function $m_0$ given $A$ (expressing one's beliefs in a context where it is only known that the truth lies in $A$), its \emph{conditional embedding} \cite{smets93b} is the least precise mass function $m$ on $\Theta$ such that $m(\cdot\vert A)=m_0$; it is obtained by transferring each mass $m_0(C)$ to $C\cup\overline{A}$, for all $C\subseteq A$. Conditional embedding is a form of ``deconditioning'', i.e., it performs the inverse of conditioning. 

\paragraph{Plausibility-probability transformation} Once a mass function representing the combined evidence has been computed, it is often used to make a decision. Decision-making methods in DST are reviewed in \cite{denoeux2019decision}. Here, we will use the simplest method \cite{cobb06}, which consists in computing a probability distribution on $\Theta$ by  normalizing the plausibilities of the singletons,
\begin{equation}
\label{eq:plp}
\forall \theta\in\Theta, \quad p(\theta)=\frac{pl(\theta)}{\sum_{k=1}^K pl(\theta_k)}.
\end{equation}
Once probabilities have been computed, a decision can be made by maximizing the expected utility. We note that this method fits well with Dempster's rule, as the plausibility of the singletons can be easily computed from \eqref{eq:prodpl} without computing the whole combined mass function. %Another way to obtain a probability distribution from a mass function, proposed by Smets \cite{smets94a}, is the pignistic transformation defined as follows:
%\begin{equation}
%    \label{eq:betp}
%\forall \theta\in\Theta, \quad p_{bet}%(\theta)=\sum_{A\subseteq\Theta: \theta\in A} \frac{m(A)}{|A|}.
%\end{equation}
%This transformation will be used in the ablation study presented in Section \ref{subsec:abla}.

\subsection{\new{Modeling the reliability of evidence}} 
\label{subsec: discount}
In the DST framework, the reliability of a source of information can be taken into account using the \emph{discounting} operation, which transforms a mass function into a weaker, less informative one and thus allows us to combine information from unreliable sources \cite{shafer1976mathematical}. Let $m$ be a mass function on $\Theta$ and $\beta$ a real number in $[0,1]$ \new{interpreted as the degree of belief that the source mass function $m$ is reliable}. The discounting operation \cite{shafer1976mathematical} with  discount rate $1-\beta$ transforms mass function $m$ into a less informative one \new{$^\beta m$ defined as a weighted sum of $m$ and the vacuous mass function $m_?$, with coefficients $\beta$ and $1-\beta$:}
\begin{equation}
^\beta m=\beta \, m +(1-\beta) \,m_?.
\label{eq:disc}
\end{equation}
\new{In the rest of this paper, we will refer to $\beta$ as a \textit{reliability coefficient}.} When $\beta=1$, we accept the mass function $m$ provided by the source and take it as a description of our knowledge; when $\beta=0$, we reject it and are left with the vacuous mass function $m_?$. 

The discounting operation \new{plays an important role in many applications of DST, where it makes it possible to take into account ``meta-knowledge'' about the reliability of a source of information.} It can be justified as follows \cite{smets94a}. Assume that $m$ is provided by a source that may be reliable ($R$) or not ($\neg R$). If the source is reliable, we adopt its opinion as ours, i.e., we set $m(\cdot\vert R)=m$. If it is not reliable, then it leaves us in a state of total ignorance, i.e., $m(\cdot\vert \neg R)=m_?$. Furthermore, assume that we have the following mass function on $\calR=\{R, \neg R\}$: $m_\calR(\{R\})=\beta$ and $m_\calR(\calR)=1-\beta$, i.e., our degree of belief that the source is reliable is equal to $\beta$. Then, combining the conditional embedding of $m(\cdot\vert R)$ with $m_\calR$ yields precisely $^\beta m$ in (\ref{eq:disc}), after marginalizing on $\Theta$.

\paragraph{Contextual discounting} %\revision{In clinical tasks, there is a consensus that the information obtained from different sources is not equally informative across diseases. Therefore, it is necessary to account for "meta-knowledge" about the reliability of a source of information in different contexts (diseases). \emph{Contextual discounting}, the generation of discounting operation, makes it possible to account for richer metaknowledge about the reliability of a source in different contexts, i.e., conditionally on different hypotheses regarding the variable of interest \cite{mercier2008refined}.}
%In the corresponding refined model, $m(\cdot\vert R)$ and $m(\cdot\vert \neg R)$ are defined as before, but our beliefs about the reliability of the source are now defined \new{in more detailed conditions} by $K$ coefficients $\beta_1,\ldots,\beta_K$, one for each condition in $\Theta$. More specifically, we have $K$ conditional mass functions defined by $m_\calR(\{R\}\vert \theta_k)=\beta_k$ and $m_\calR(\calR\vert \theta_k)=1-\beta_k$, for $k=1,\ldots, K$. In this model, $\beta_k$ is, thus, the degree of belief that the source of information is reliable, given that the true state is $\theta_k$. 

In \cite{mercier2008refined}, the authors generalize the discounting operation using the notion of \emph{contextual discounting}, which makes it possible to account for richer metaknowledge about the reliability of a source \new{in different contexts, i.e., conditionally on different hypotheses regarding the variable of interest.}
In the corresponding refined model, $m(\cdot\vert R)$ and $m(\cdot\vert \neg R)$ are defined as before, but our beliefs about the reliability of the source are now defined by $K$ coefficients $\beta_1,\ldots,\beta_K$, one for each state in $\Theta$. More specifically, we have $K$ conditional mass functions defined by $m_\calR(\{R\}\vert \theta_k)=\beta_k$ and $m_\calR(\calR\vert \theta_k)=1-\beta_k$, for $k=1,\ldots, K$. In this model, $\beta_k$ is, thus, the degree of belief that the source of information is reliable, given that the true state is $\theta_k$. \revision{As shown in \cite{mercier2008refined},} combining the conditional embeddings of $m(\cdot\vert R)$ and $m_\calR(\cdot \vert \theta_k)$ for $k=1, \ldots, K$ by Dempster's rule yields the following discounted mass function,
\begin{equation} 
\label{eq:cdisc}
^{\bbeta} m(A) = \sum_{B \subseteq A} m(B) \left(\prod_{\theta_k \in A \setminus B}(1-\beta_k) \prod_{\theta_l \in \overline{A}}\beta_l\right) 
\end{equation}
for all $A \subseteq \Theta$, where $\bbeta=(\beta_1,\ldots,\beta_K)$ is the vector of all reliability coefficients, and a product of terms is equal to 1 if the index set is empty. \new{In many applications, we actually do not need to compute the whole mass function \eqref{eq:cdisc}: we can compute only the associated contour function $^{\bbeta}pl$, which is all we need for decision-making.} As shown in \cite{mercier2008refined}, this contour function is equal to  
\begin{equation} 
\label{eq:cdisc_contour}
%\begin{align}
^{\bbeta}pl(\theta_k) =1-\beta_k + \beta_k pl(\theta_k) , \quad k=1, \ldots, K.
%\end{align}
\end{equation}
\new{It can be computed in linear time with respect to the size of $\Theta$, instead of exponential time for $^{\bbeta} m$}. An evidential $k$ nearest neighbor rule based on the contextual discounting operation was introduced in \cite{denoeux19f}. % \revision{Figure \ref{fig: contextual} shows an example of contextual discounting operations based on the corresponding contour function. For each given contour function $pl(\theta_k)$, the discounting is applied under two conditions:
%\begin{enumerate}
%    \item if there is a reliability indicator $\beta_k$ given that the true state is \textbf{in} $\theta_k$, the contour function $pl(\theta_k)$ is discounted to $\beta_k pl(\theta_k)$;
%    \item otherwise, the reliability indicator $1-\beta_k$ given that the true state is \textbf{NOT} in $\theta_k$, the contour function $pl(\theta_k)$ discounted to $(1-\beta_k) \times pl(\Theta)$, i.e., $1-\beta_k$.
%\end{enumerate}
%\begin{figure}
%    \centering
%    \includegraphics[width=\textwidth]{figure/Contextual_discounting.pdf}
%    \caption{\revision{Example of contextual discounting operation of contour function $pl(\theta_k)$ on contextual space $\{ \theta_1, \ldots, \theta_k\}$, with related reliability coefficient $\{ \beta_1, \ldots, \beta_k\}$.} }
%    \label{fig: contextual}
%\end{figure}}

\revision{
\begin{Ex}
Consider a simplified diagnostic problem in which a patient may have one of two diseases denoted by $\theta_1$ and $\theta_2$. Assume that $\theta_1$ is a heart disease while $\theta_2$ is a lung disease. A cardiologist examines the patient and  describes his opinion by the following mass function on $\Theta=\{\theta_1,\theta_2\}$: $m(\{\theta_1\})=0.7$, $m(\{\theta_2\})=0.2$, $m(\Theta)=0.1$, i.e., his degrees of belief in $\theta_1$ and $\theta_2$ are, respectively, 0.7 and 0.2. Furthermore, suppose that the cardiologist is fully reliable to diagnose heart diseases ($\beta_1=1$), i.e., if the true state of the patient is $\theta_1$, the physician's opinion can be fully trusted, whereas he is only 60\% reliable to diagnose lung diseases ($\beta_2=0.6$), i.e., if $\theta_2$ is the true disease, there only is only 60\% chance that the physician's diagnostic is relevant. Applying formula \eqref{eq:cdisc} to $m$ gives the following discounted mass function:
\begin{align*}
^{\bbeta} m(\{\theta_1\}) &= \beta_2 m(\{\theta_1\})=0.42\\
^{\bbeta} m(\{\theta_2\}) &= \beta_1 m(\{\theta_2\})=0.2\\
^{\bbeta} m(\Theta) &= 1- \beta_2 m(\{\theta_1\})-\beta_1 m(\{\theta_2\})=0.38.
\end{align*}
The contour function of the original mass function $m$ is
\begin{align*}
pl(\{\theta_1\}) = 0.7+0.1=0.8\\
pl(\{\theta_2\}) = 0.2+0.1=0.3.
\end{align*}
After contextual discounting, we get
\begin{align*}
^{\bbeta} pl(\{\theta_1\}) = 0.42+0.38=0.8\\
^{\bbeta} pl(\{\theta_2\}) = 0.2+0.38=0.58.
\end{align*}
 We can check that $^{\bbeta} pl(\{\theta_1\}) =  1-1 + 1\times0.8$ and $^{\bbeta} pl(\{\theta_2\}) = 1-0.6+0.6\times 0.3$, which is consistent with \eqref{eq:cdisc_contour}.

\end{Ex}
}

\subsection{Evidential neural network}
\label{subsec: enn}
\begin{figure}
    \centering
    \includegraphics[width=0.8\textwidth]{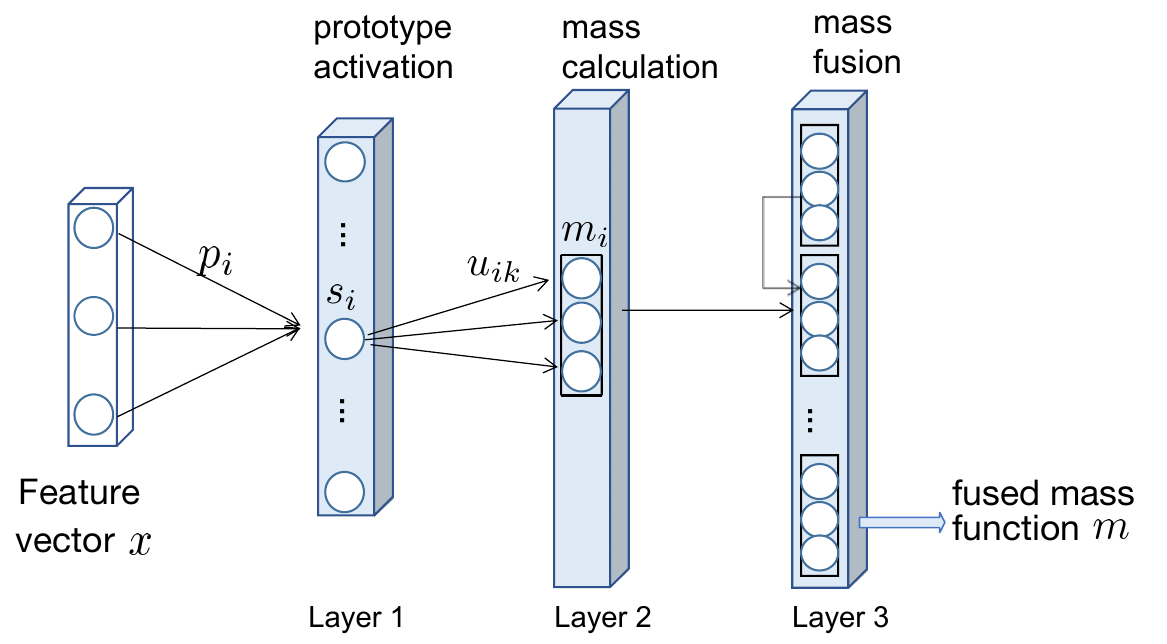}
    \caption{The evidential neural network model.}
    \label{fig: enn}
\end{figure}

In \cite{denoeux2000neural}, Den{\oe}ux proposed a DST-based evidential neural network (ENN) classifier in which mass functions are computed based on distances between the input vector and prototypes. As shown in Figure \ref{fig: enn}, the ENN model comprises a prototype activation layer, a mass calculation layer, and a combination layer.

The prototype activation layer comprises $I$ units, whose weight vectors are prototypes $\bp_1,\ldots, \bp_I$ in input space. The activation of unit $i$ in the prototype layer is
\begin{equation}
\label{eq:enn1}
    s_i=\alpha _i \exp(-\gamma_i \left \| \bx-\bp_i \right \| ^2),
\end{equation}
where $\gamma_i>0$ and $\alpha_i \in [0,1]$ are two parameters. Each quantity $s_i$ can be interpreted as a degree of similarity between input vector $\bx$ and prototype $\bp_i$.

The second hidden layer computes mass functions $m_i$ representing the evidence of each prototype $\bp_i$, using the following equations: 
\begin{subequations}
\label{eq:enn2}
    \begin{align}
m_i(\{\theta _{k}\})&=u_{ik}s_i, \quad k=1,\ldots, K,\\
m_{i}(\Theta)&=1-s_i, 
\end{align}
\end{subequations}
where $u_{ik}$ is the membership degree of prototype $i$ to class $\theta_k$, and $\sum _{k=1}^K u_{ik}=1$. The mass function $m_i$ can thus be seen as a discounted Bayesian mass function, with a discount rate $1-s_i$; its focal sets are singletons and $\Theta$. The mass assigned to $\Theta$ increases with the distance between $\bx$ and $\bp_i$. Finally, the third layer combines the $I$ mass functions $m_1,\ldots,m_I$ using Dempster's rule \eqref{eq:demp1}. The output mass function $m=\bigoplus_{i=1}^I m_i$ is a discounted Bayesian mass function that summarizes the evidence of the $I$ prototypes.

The idea of applying the above model to features extracted by a convolutional neural network (CNN) was first proposed by Tong et al. in \cite{tong21b}. In this approach, the ENN module becomes an ``evidential layer'', which is plugged into the output of a CNN instead of the usual softmax layer. The feature extraction and evidential modules are trained simultaneously. Huang et al. applied the ENN model to medical image segmentation within a deep evidential segmentation network \cite{huang2022lymphoma}.

\begin{Rem}
    The approach described in this section should not be confused with the ``evidential deep learning'' approach introduced in \cite{sensoy18} and applied to brain tumor segmentation in \cite{zou2022tbrats}. The latter approach is based on learning the parameters of a Dirichlet distribution that represents second-order uncertainty on the class probabilities. Although the parameters of the Dirichlet distribution can be formally identified to a mass function whose focal sets are the singletons $\{\theta_k\}$ and the whole frame $\Theta$, this is actually a Bayesian approach that learns a probability distribution over the class probabilities through a suitable loss function.
\end{Rem}

\subsection{Multimodal medical image fusion}
\label{subsec:multimod}

\revision{Multimodal medical image fusion can be performed at the pixel, feature or decision level. Pixel-level fusion  is the traditional approach; it can be conducted directly in the spatial domain or indirectly through the application of transformations and representations. The fusion of high-level features is typically performed by a neural network  learning a shared representation or a joint embedding space derived from multimodal features. Decision fusion consists in pooling decisions made independently from different image modalities; it can be performed  with  traditional or deep-learning approaches. In the following, we review previous work on multimodal medical image fusion, emphasizing the distinction between traditional and deep-learning approaches.

\subsubsection{Traditional approaches}
Traditional fusion methods aim at combining  relevant information (either pixels themselves or low-level image features) from multiple images to produce a single fused image with enhanced features for further analysis. Four main approaches have been proposed: multi-scale transformation,  sparse representation extraction, edge-preserving filters, and meta-heuristic optimization. The first three approaches focus on  effective image representation, while the last one aims at combining the represented features efficiently. 

The multi-scale transform approach decomposes images into different scales or frequency components using techniques such as wavelet transform \cite{singh2009multimodal}, contourlet transforms \cite{yang2008multimodality}, pyramid transforms \cite{du2016union} or curvelet transform \cite{arif2020fast}, allowing relevant features from each source image to be combined. Sparse representation extraction assumes that multimodal images can be represented as a sparse linear combination of basis functions; search techniques such as dictionary learning \cite{kim2016joint} or sparse coding with dictionary learning \cite{veshki2022coupled} are used to obtain the sparse image representation and to merge images focusing on the most important features. Edge-preserving filters ensure the preservation of edges while smoothing images to ensure the fusion of critical features without blurring \cite{tan2021multi}. Commonly used filters include bilateral filters \cite{li2021multimodal}, guided filters \cite{pei2020two}, anisotropic diffusion \cite{wang2015robust}, and total variation minimization \cite{zhao2017medical}. The three above approaches can be used independently or in combination, which often yields better results. For example, in \cite{hu2020multi}, Hu et al. propose a multimodal medical image fusion method based on separable dictionary learning and Gabor filtering; in \cite{wang2020multi}, Wang et al. describe a multimodal medical image fusion method using Laplacian pyramid and adaptive sparse representations; in \cite{liu2015general}, Liu et al. introduce a general image fusion framework based on multi-scale transform and sparse representation. 

In addition to studying effective image representations, a complementary research direction has been to  design meta-heuristic  optimization algorithms allowing one to find the best fusion parameters for combining features obtained by different transform, sparse or fitting algorithms. Many approaches use meta-heuristic optimization techniques such as genetic algorithms \cite{arif2020fast}, particle swarm optimisation \cite{tannaz2020fusion} or ant colony optimisation \cite{shahdoosti2019mri}. %However, the outcomes of the above-mentioned traditional approaches are still unsatisfying due to the limited information representation and fusion optimization ability.

\subsubsection{Deep learning approaches}

Recent advances in deep learning have allowed breakthroughs in medical image fusion by making it to learn a joint embedding or a shared representation space from multiple features. Recent techniques include  adversarial learning \cite{safari2023medfusiongan}, co-training \cite{zhang2023multi}, multi-kernel learning \cite{meng2022feature}, multi-task learning \cite{liu2022sf}, etc. These methods exploit the ability of neural networks to extract meaningful representations and perform fusion in high-level feature spaces with learnable feature fusion rules. These approaches enable more sophisticated and robust image fusion, capable of handling complex relationships and producing high-quality fused image features. Here, we summarize three important models  commonly used for multi-model medical image fusion.

\paragraph{Convolutional Neural Networks}
Convolutional neural networks (CNNs) are widely used in image processing due to their strong feature representation capability. Within CNNs, various fusion operations can be used to effectively integrate information from different imaging modalities. Such operations include but are not limited to, concatenation, element-wise addition and multiplication, weighted sum, max pooling, etc. Fusion can occur at different stages of the network, i.e., early, middle, or late stages. 

Early fusion stacks different modalities along a channel dimension and feeds into a single CNN \cite{liang2019mcfnet}. This is the simplest operation but it requires high image registration quality. In the case of middle fusion, separate CNN branches are employed to extract features from each modality, which are subsequently concatenated at the feature level or fused in a particular common representation space. More recently, transformer-based CNN architectures, such as the Vision Transformer (ViT) \cite{han2022survey}, have also demonstrated considerable versatility in handling diverse types of data with the introduction of an attention mechanism \cite{li2024crossfuse}. CNNs can also be integrated with some traditional fusion ideas to obtain more robust fusion results using, e.g., the multiscale transformer \cite{tang2022matr} or multiscale residual pyramid attention network \cite{fu2021multiscale}.

In contrast to the emphasis on image pixels or features in earlier fusion techniques, later fusion places greater importance on the aggregation of high-level decisions. It integrates information derived from preliminary classifications with the application of appropriate fusion rules. Approaches can be classified into two main categories: 1) \textit{hard fusion} methods, which merge logical information membership values, such as model ensembling with majority or average voting \cite{khan2024hybrid}; and 2) \textit{soft fusion} methods, where classifiers assign numerical values to reflect their confidence in decisions,  as exemplified by fuzzy voting \cite{hall2008framework, foo2013high}. 

\paragraph{Encoder-Decoder Networks}
Encoder-decoder networks are another type of convolutional neural network  commonly used for image segmentation and reconstruction. Within the encoder-decoder network, multiple encoders are used to extract deep features  from each modality. These features are subsequently integrated either through a straightforward concatenation process or through  a latent layer or learnt joint embedding space. The fused features are then passed to the decoder to produce the final image. Compared with CNNs, the Encoder-decoder architecture offers a more structured and effective fusion framework with enhanced feature representation, precise spatial alignment, and flexible and effective fusion strategies. Multimodal Transformer (MMT) \cite{xu2023multimodal} is one of the most sophisticated forms of multimodal encoder-decoder networks that employ self-attention mechanisms to integrate and process multimodal data in an effective manner; nnFormer \cite{zhou2023nnformer} has been identified as the most advanced model for multimodal MRI brain tumor segmentation. 

\paragraph{Generative Adversarial Networks}

Generative Adversarial Networks (GANs), composed of a generator and a discriminator, are capable of learning complex relationships between disparate modalities through the generation of highly realistic images via unsupervised adversarial training \cite{goodfellow14}. In the context of multimodal medical image fusion, the generator learns to generate a fused image that combines the semantic features of the inputs from different modalities. The discriminator guides the generator to produce high-quality fused images by distinguishing between the fused and the real images. GAN-based fusion methods are particularly useful for advanced medical image fusion tasks where the quality and realism of the fused image are of paramount importance, such as the combination of structural and functional imaging modalities. For example, in \cite{zhang2023transformer}, the authors propose a conditional generative adversarial network with a transformer for multimodal image fusion by introducing a wavelet fusion module to maintain long-distance dependencies across domains; in \cite{safari2023medfusiongan}, the authors introduce an unsupervised medical fusion generative adversarial network to generate an image with CT bone structure and MRI soft tissue contrast by fusing CT and MRI image sequences. 
}

Although a lot of research has been devoted to the study of multimodal medical image segmentation and promising experimental results have been obtained, modeling the reliability of each modality in a given context and quantifying the uncertainty on the outcome of the fusion process remain challenging research questions. In this paper, we address these questions using a deep evidential fusion framework combining deep learning with DST, and taking into account the reliability of each of the modalities being combined. The proposed decision-fusion framework is described in detail in the following section. 

\section{Proposed framework}
\label{sec: proposed}

The main idea of this paper is to hybridize a deep evidential fusion framework with uncertainty quantification and reliability learning for multimodal medical image segmentation under the framework of DST.  The architecture of the system is described in Section \ref{subsec:archi}, and the loss function used to train the whole framework end-to-end is presented in Section \ref{subsec:loss}.

\subsection{Architecture}
\label{subsec:archi}

\begin{figure}
\centering
\includegraphics[width=\textwidth]{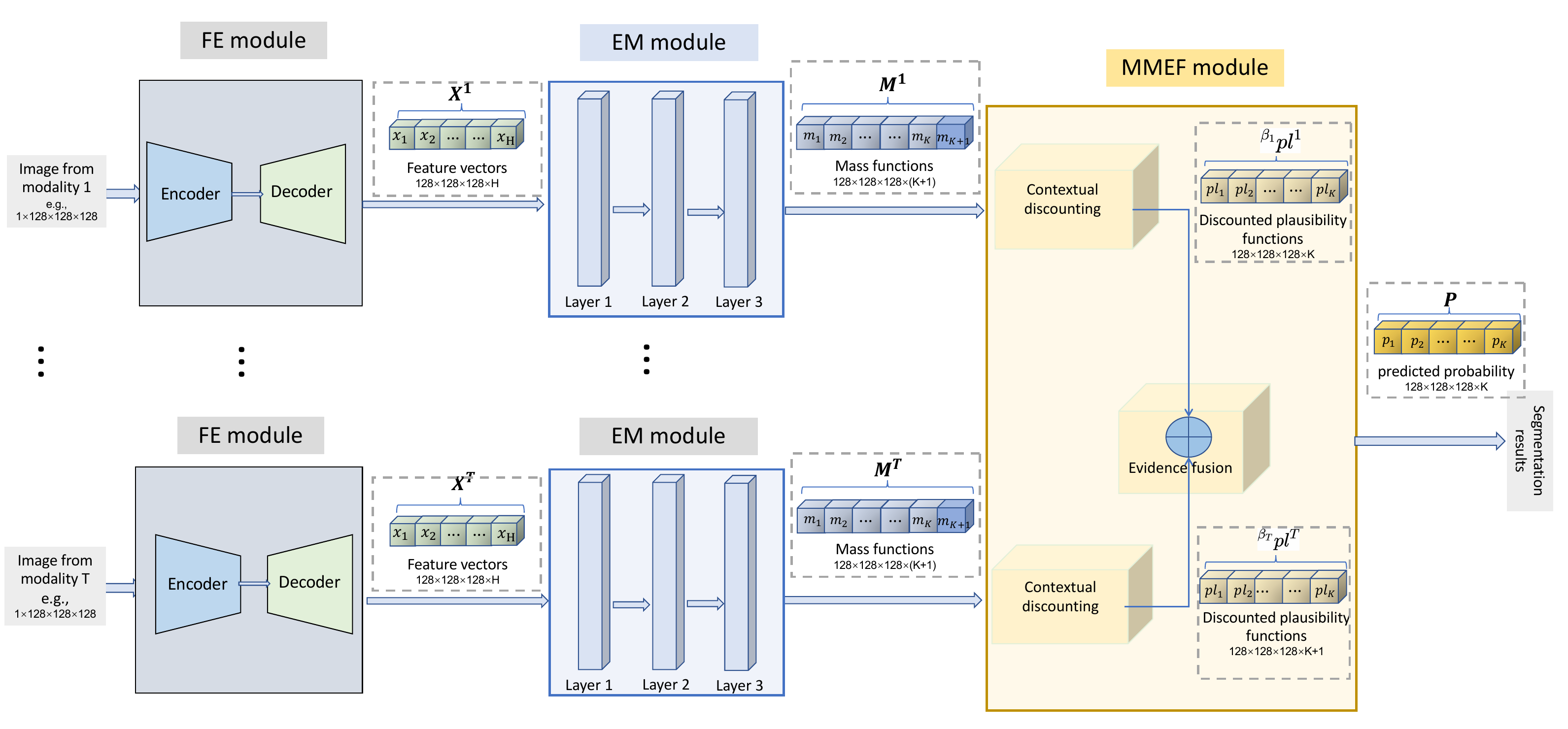}
\caption{The proposed deep evidential fusion framework. It is composed of encoder-decoder \emph{feature extraction (FE)} modules that represent images using deep features, \emph{evidence mapping (EM)} modules that map deep features into mass functions, and a \emph{multimodal evidence fusion (MMEF)} module that combines evidence from different modalities.}
\label{fig:architecture}
\end{figure}

The proposed framework is depicted in Figure \ref{fig:architecture}. Features are first extracted from different modalities using independent encoder-decoder feature-extraction (FE) modules. The features from each modality are then transformed into mass functions using evidence mapping (EM) modules. Finally, mass functions are discounted and combined in a multi-modality evidence fusion (MMEF) module. These modules are described in greater detail below.

\subsubsection{Feature-extraction (FE) module} 
\begin{figure}
\includegraphics[width=\textwidth]{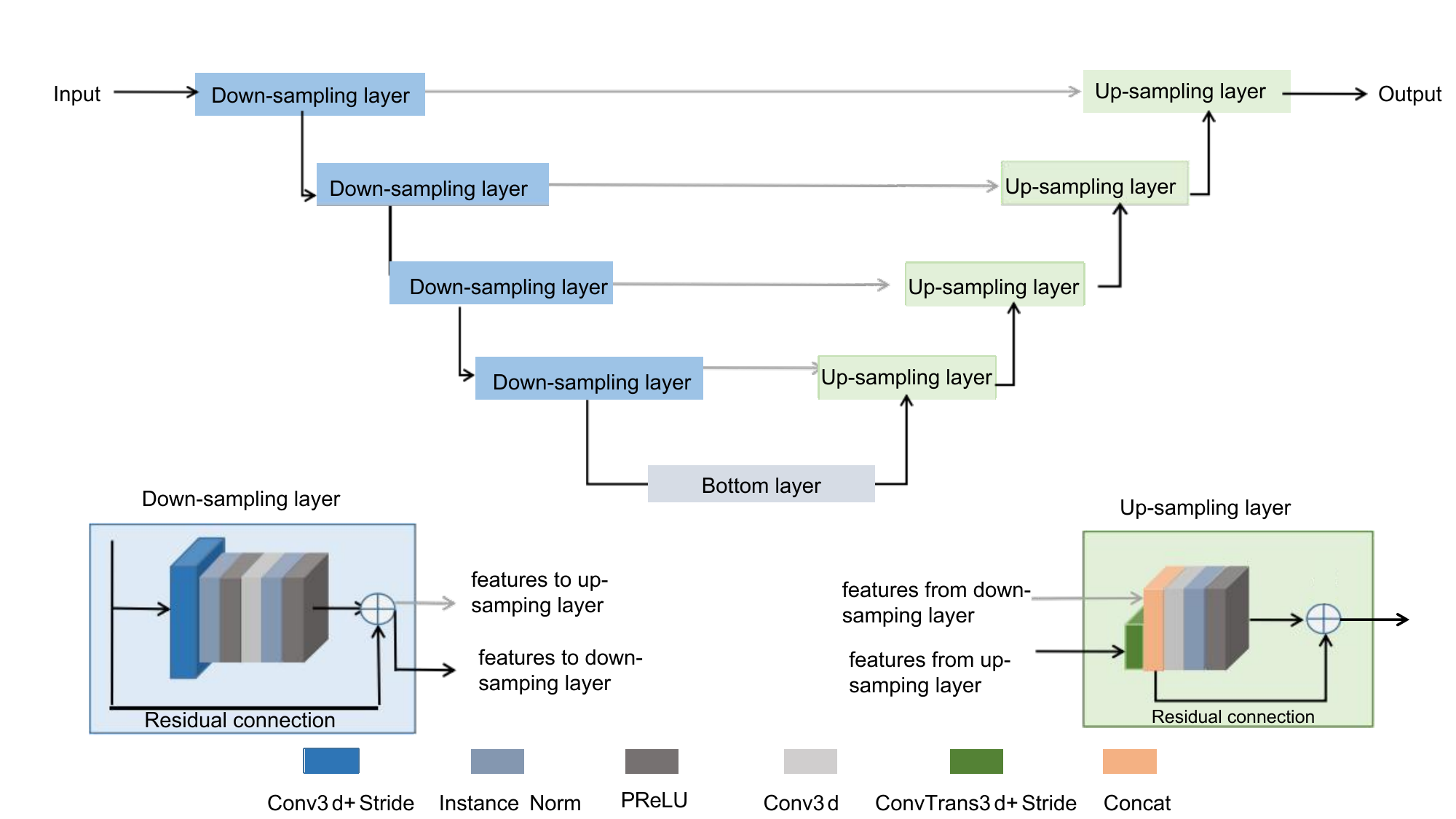}
\caption{\new{Schematic description of a UNet-based FE module. The network consists of a contracting path (down-sampling layers) and an expansive path (up-sampling layers), which gives it the u-shaped architecture.} Reproduced based on \cite{kerfoot2018left}.}
\label{fig: FE-unet}
\end{figure}

Deep neural network architectures have been shown to be very powerful for extracting relevant information from high-dimensional data. Our approach is compatible with any deep FE architecture. The baseline model considered in this paper is  UNet \cite{kerfoot2018left}, a foundational medical image segmentation model.
%The number of filters was set as $(8, 16, 32, 64, 128)$ with kernel size equal to five and convolutional strides equal to $(2, 2, 2, 2)$ for layers from left to right. Our framework could, alternatively, incorporate any state-of-the-art feature extraction model such as, e.g., nnUNet \cite{isensee2018nnu}. %\textbf{For nnUNet, the kernel size was set as $(3, (1,1, 3), 3, 3)$ and the upsample kernel size was set as $(2,2,1)$ with strides $((1,1,1), 2, 2, 1)$.} %Comparison and analysis of different feature extraction modules will be given in Section \ref{subsubsec: acc-brats}.
As illustrated in Figure \ref{fig: FE-unet}, a UNet-based feature extraction module incorporates residual connections within each layer, following the same architecture as in \cite{huang2022lymphoma}. Each layer of the module comprises encoding and decoding paths, connected by skip connections. In the encoding path (represented by blue blocks), the data undergoes downsampling through stride convolutions, while the decoding path (represented by green blocks) employs stride transpose convolutions for upsampling. The bottom layer, represented by the gray block, serves as the base connection without performing any down or up-sampling of the data. \new{In Section \ref{subsec: brain}, in addition to UNet, we will also consider the more recent nnUNet \cite{isensee2018nnu} and nnFormer \cite{zhou2023nnformer} models as alternative FE modules. The settings of these modules will be described in Section \ref{subsec: setting}.}

\subsubsection{Evidence mapping (EM) module}
\label{subsubsec:ES}

The EM module is based on the ENN architecture recalled in Section \ref{subsec: enn}. It is identical to that described in \cite{huang2022lymphoma}. As illustrated in Figure \ref{fig:architecture}, we have one such module for each modality. The input to each module is a tensor containing the $H$ features extracted for each voxel. The prototypes are, thus, vectors in the $H$-dimensional space of features extracted from modality $t$ images by the FE module. \new{As explained in Section \ref{subsec: enn}, a prototype layer first computes the similarities between feature vectors and prototypes using \eqref{eq:enn1}.  The next layer computes mass functions for each prototype using \eqref{eq:enn2} (see Figure \ref{fig: enn}). Finally, the prototype-based mass functions are combined by Dempster's rule  \eqref{eq:demp1} in a third layer.} Denoting by $\Theta=\{\theta_1,\ldots,\theta_K\}$ the set of classes, the EM module thus computes, for each voxel $n$ and modality $t$, a mass function\footnote{\new{Throughout this paper, we use an upper index $t$ to denote modalities, and lower indices $n$ and $k$ to denote, respectively,  voxels and classes.}} $m^t_n$ with focal sets $\{\theta_k\}$, $k=1,\ldots,K$ and $\Theta$. The mass $m^t_n(\Theta)$ is a measure of the segmentation uncertainty for classifying voxel $n$ in the image of modality $t$.

\subsubsection{Multi-modality evidence fusion (MMEF) module}
\label{subsubsec:MMEF}

\new{This module first transforms the contour functions from the EM modules using the contextual discounting operation recalled in Section \ref{subsec: discount}. The contour function for voxel $n$ and modality $t$ is obtained from mass function $m_n^t$ as 
\begin{equation*}
   pl_n^t(\theta_k) =  m_n^t(\{\theta_k\})+m_n^t(\Theta), \quad k=1,\ldots,K.
\end{equation*}
Using \eqref{eq:cdisc_contour}, the discounted contour function is given by
\begin{equation}
    {^{\bbeta^t}} pl_n^t(\theta_k) = 1-\beta^t_k+ \beta^t_k pl_n^t(\theta_k)  , \quad k=1,\ldots,K,
\label{eq:c_discounting_t}
\end{equation}
where $\bbeta^t=(\beta^t_1,\ldots,\beta^t_K)$ is the vector of discounting (reliability) coefficients for modality $t$. We recall that $\beta^t_k$ represents our degree of belief that the modality $t$ is reliable when it is known that the actual class of voxel $n$ is $\theta_k$. From \eqref{eq:prodpl}, the combined contour function at voxel $n$ can then be computed up to a multiplicative constant  by multiplying the contour functions for the $T$ modalities as
\[
{^\bbeta}pl_n(\theta_k)  \propto \prod_{t=1}^T {^{\bbeta^t}}pl^t_n(\theta_k), \quad k=1,\ldots,K,
\]
where $\bbeta=(\bbeta^1,\ldots\bbeta^T)$ is the vector of $KT$ reliability coefficients for the the $K$ classes and $T$ modalities. Finally,  the  predicted probability distribution $P_n$ for voxel $n$ after combining evidence from the $T$ modalities is obtained from \eqref{eq:plp} as
\begin{equation}
      {^\bbeta}p_n(\theta_k)= \frac{{^\bbeta}pl_n(\theta_k)}{\sum_{l=1}^K {^\bbeta}pl_n(\theta_l)}=
      \frac {\prod_{t=1}^{T} \left(1-\beta^t_k+ \beta^t_k pl_n^t(\theta_k)  \right)} {\sum_{l=1}^{K} \prod_{t=1}^{T}  \left(1-\beta^t_l+ \beta^t_l pl_n^t(\theta_l)  \right)} 
      \label{eq:dis_pl}, \quad k=1,\ldots,K.
\end{equation}
The learnable parameters in this module are the $KT$ reliability coefficients in vector $\bbeta$.}

\subsection{Loss function}
\label{subsec:loss}
 
The whole framework  is optimized by minimizing the following loss function, 
\begin{equation*}
%\label{eq:loss}
\loss=\loss_{s}+\loss_{f},
\end{equation*}
where
\begin{itemize}
    \item The term $\loss_{s}$ is the Dice loss quantifying the segmentation performance of each source modality independently, with
\begin{equation}
    \loss_{s}= \sum_{t=1}^{T} \left[
    1-\frac{2 \sum_{n=1}^{N}\sum_{k=1}^{K} m_{n}^t(\{\theta_k\}) \times G_{kn}} {\sum_{n=1}^{N}\sum_{k=1}^{K} (m_{n}^t(\{\theta_k\}) + G_{kn})} \right],
%\label{eq:loss_s}
\end{equation}
where $N$ is the number of voxels, and $G_{kn}=1$ if voxel $n$ belongs to class $\theta_k$, and $G_{kn}=0$ otherwise; 
\item The  term $\loss_{f}$ quantifies the segmentation performance after combination:
\begin{equation}
    \loss_{f}=1-\frac{2 \sum_{n=1}^{N}\sum_{k=1}^{K} {^\bbeta}p_{n}(\theta_k) \times G_{kn}} {\sum_{n=1}^{N}\sum_{k=1}^{K} {^\bbeta}p_{n}(\theta_k) + G_{kn}},
%\label{eq:loss}
\end{equation}
where ${^\bbeta}p_{n}$ is the predicted probability distribution for voxel $n$ given by  \eqref{eq:dis_pl}. 
\end{itemize}

The learnable parameters are the weights of the FE module, the prototypes and associated parameters $\alpha_i$, $\gamma_i$ and $u_{ik}$ of the EM module, and the reliability coefficients $\beta^t_k$ in the MMEF module. Learning the reliability coefficients is an original feature of our approach. As shown in Sections \ref{subsec: lym} and \ref{subsec: brain}, these coefficients can allow us to gain some insight into the multi-modality segmentation process. 

\section{Experiments and results}
\label{sec: exp}

In this section, the proposed framework described in Section \ref{sec: proposed} is applied to two real multimodal medical image datasets. The experimental settings are first described in Section \ref{subsec: setting}. The results on the two datasets are then reported in Sections \ref{subsec: lym} and \ref{subsec: brain}. 

\subsection{Experimental settings}
\label{subsec: setting}
\paragraph{Datasets}
%\label{subsec:data}
The proposed framework was tested on two multimodal medical image datasets. 

\begin{figure}
\centering
\includegraphics[width=\textwidth]{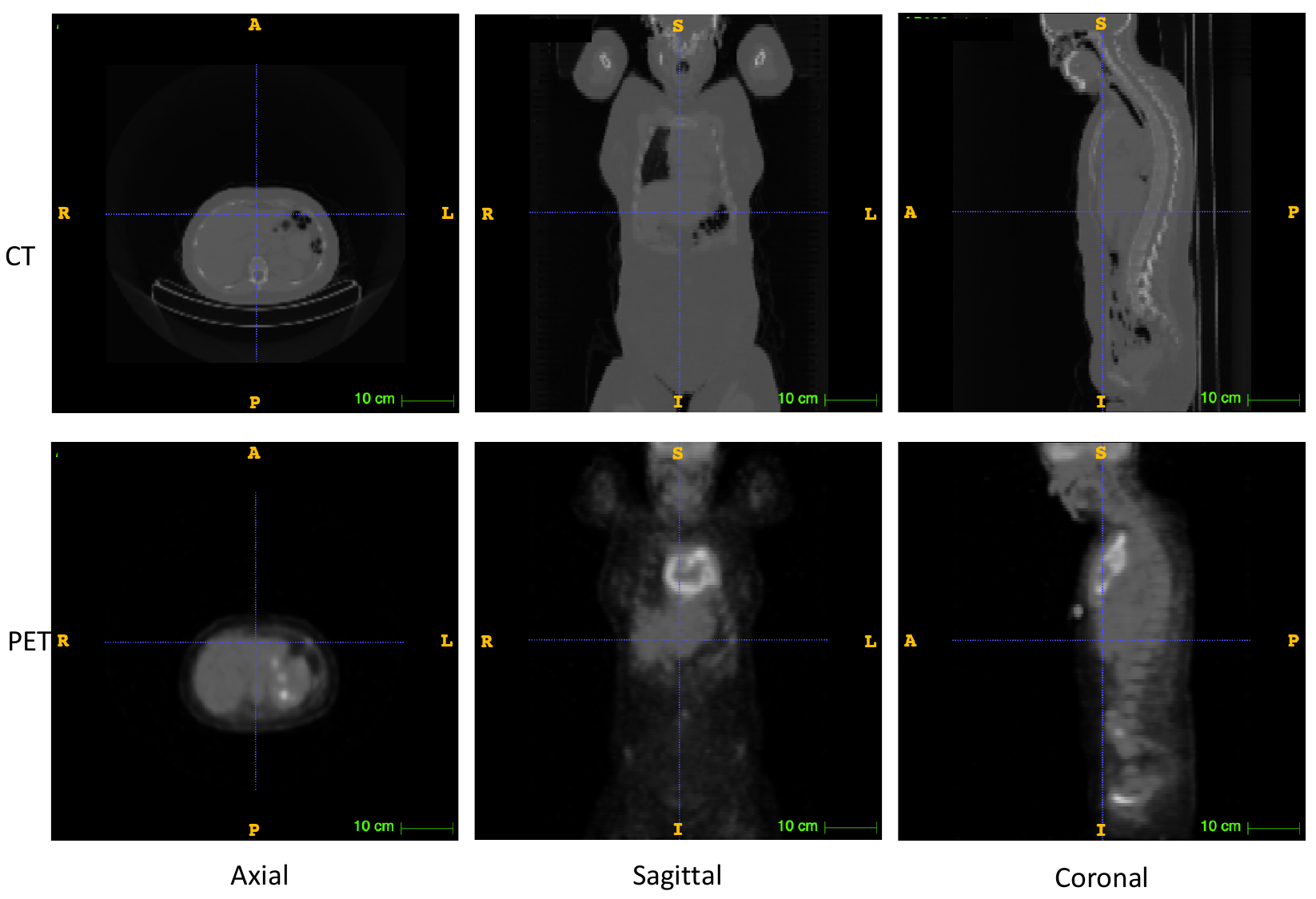}
\caption{\new{Example of a patient with lymphomas. The first and second rows showcase, respectively, CT and PET slices, depicting axial, sagittal, and coronal views. The lymphomas correspond to the bright regions in PET slices.}}
\label{fig:lymphoma-intensity}
\end{figure}

The \emph{PET-CT lymphoma dataset} contains 3D images from 173 patients who were diagnosed with large B-cell lymphomas and underwent PET-CT examination\footnote{The study was approved as a retrospective study by the Henri Becquerel Center Institutional Review Board.}. \new{For lymphoma segmentation, PET imaging helps identify active tumor sites by highlighting areas of increased metabolic activity. In contrast, CT imaging provides anatomical information about the size, shape, location, and surrounding structures of lymphoma tumors. While PET image makes it possible to obtain functional information about the tumor and surrounding tissues, CT images provide complementary anatomical details allowing for more accurate segmentation.} The lymphomas in mask images were delineated manually by experts and considered as ground truth.  Figure \ref{fig:lymphoma-intensity} shows an example of PET and CT images of a patient with lymphomas. The PET and CT images and the corresponding mask images have different sizes and spatial resolutions due to the use of different imaging machines and operations. For CT images, the size varies from $267\times 512\times512$ to $478\times 512\times512$. For PET images, the size varies from $276\times 144\times144$ to $407\times 256\times256$.  

\begin{figure}
\centering
\includegraphics[width=\textwidth]{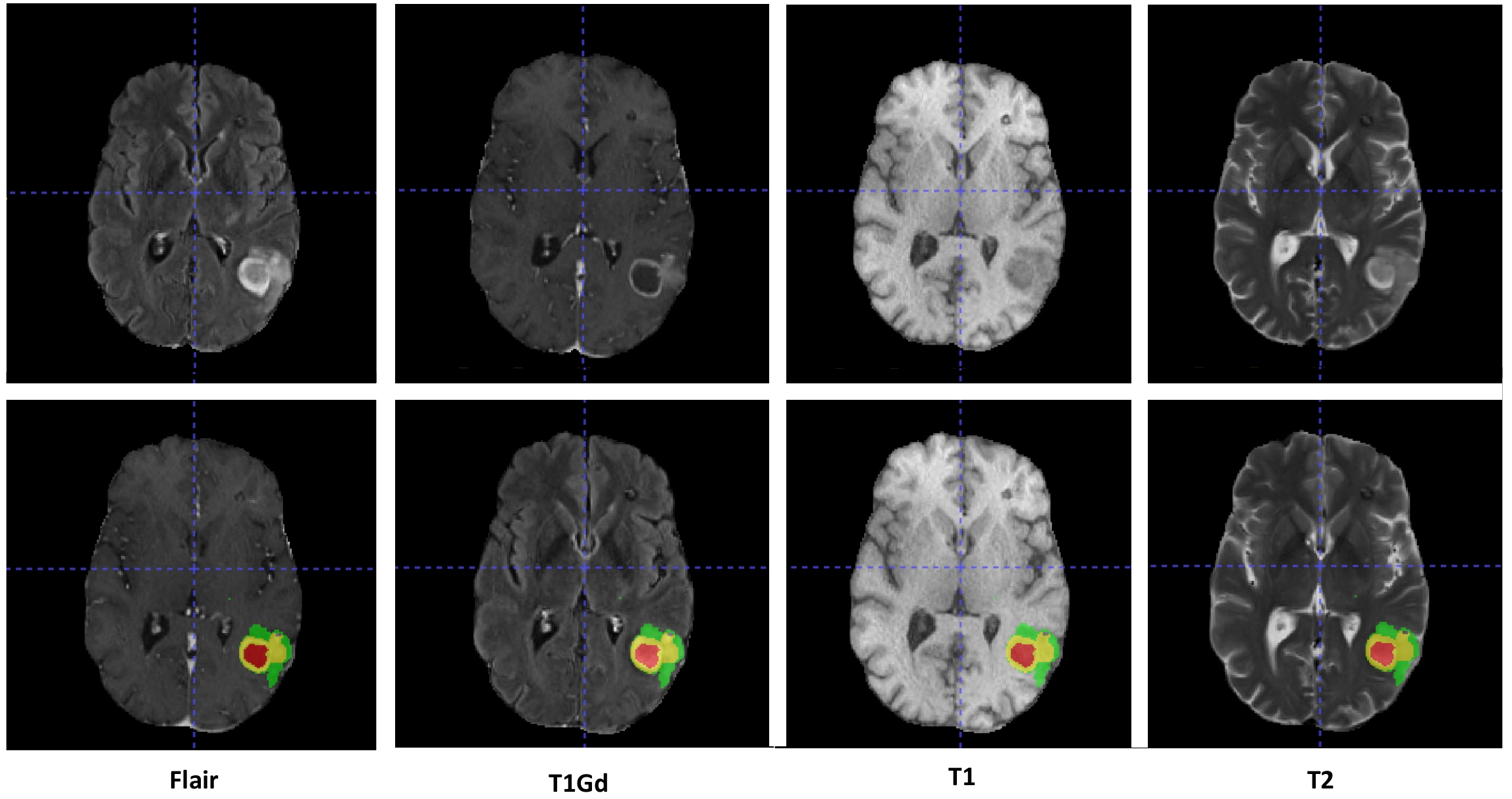}
\caption{\new{Examples of a patient with brain tumors in four MRI modalities: \new{FLAIR, T1Gd, T1, and T2}. The first and second rows show, respectively, the original images and the images with tumor masks for the three classes: peritumoral edema (ED, green), enhancing tumor (ET, yellow), and necrotic tumor core or non-enhancing tumor (NCR/NET, red).}}
\label{fig:image_tumor_intensity}
\end{figure}

The \emph{multi-MRI brain tumor} dataset was made available for the BraTS2021 challenge \cite{baid2021rsna}. The original BraTS2021 dataset comprises training, validation, and test sets with, respectively, 1251, 219, and 570 cases. There are four modalities: FLAIR, T1Gd, T1, and T2 with $240\times 240 \times 155$ voxels. Figure \ref{fig:image_tumor_intensity} shows examples of four-modality MRI slices for one patient. The appearance of brain tumors varies in different modalities \cite{baid2021rsna}. 
\new{T1Gd MRI images are obtained following the administration of a gadolinium-based contrast agent that enhances areas with disrupted blood-brain barrier such as tumor regions, making tumors appear hyperintense (bright) and improving the visibility of tumor margins. FLAIR MRI images suppress the signal from cerebrospinal fluid (CSF), highlighting pathological changes while suppressing the CSF signal. T2 MRI images are sensitive to tissue water content and provide good contrast between soft tissues. Tumors with increased water content often appear hyperintense (bright) on T2 images. T1 MRI images are crucial for identifying tumor location and structural details by their excellent anatomical detail.} Annotations of scans comprise gadolinium (GD)-enhancing tumor (ET), necrotic and non-enhancing tumor core (NRC/NET), and peritumoral edema (ED). The task of the BraTS2021 challenge was to segment the images into three overlapping regions: ET, tumor core (TC, the union of ET and NRC/NET), and whole tumor (WT, the union of ET, NRC/NET, and ED). \new{In this work, we evaluated the segmentation performances with respect to these three overlapping regions to allow a fair comparison with other state-of-the-art methods. Additionally, we also compared the results with respect to the three original non-overlapping tumor regions to highlight the impact of contextual discounting on subregion segmentation.}
%Moreover, the tumor boundaries are blurred, making it hard to delineate different tumors precisely. Annotations of scans comprise gadolinium (GD)-enhancing tumor (ET), necrotic and non-enhancing tumor core (NRC/NET), and peritumoral edema (ED).

\paragraph{Pre-processing}
For the PET-CT dataset, we first normalized the PET, CT and mask images: (1) for PET images, we applied a random intensity shift and scale to each channel with a shift value of 0 and scale value of 0.1; (2) for CT images, the shift and scale values were set to 1000 and 1/2000; (3) for mask images, the intensity value was normalized into the $[0,1]$ interval by replacing the outside value by $1$. We then resized the PET and CT images to $256\times 256\times 128$ by linear interpolation and mask images to $256\times 256\times 128$ by nearest neighbor interpolation. Lastly, CT and PET images were registered using B-spline interpolation. Following \cite{huang2022lymphoma}, we randomly divided the 173 scans into subsets of size 138, 17, and 18 for, respectively, training, validation, and test. The training process was then repeated five times to test the stability of our framework, with different data used exactly once as the validation and test data.

For the BraTS2021 dataset, we used the same pre-processing operation as in \cite{peiris2021volumetric}. We first performed a min-max scaling operation and clipped intensity values to standardize all volumes; we then cropped/padded the volumes to a fixed size of $128 \times 128 \times 128$ by removing the unnecessary background (the cropping/padding operation was only applied to training data). No data augmentation technique was applied, and no additional data was used in this study. Since the ground truth labels are unavailable for the validation and test sets, we trained and tested our framework with the training set. Following \cite{peiris2021volumetric}, we randomly divided the 1251 training scans into subsets of 834, 208, and 209 cases for training, validation, and testing, respectively. The process was repeated five times to test the stability of our framework. All the preprocessing methods mentioned in this paper can be found in the SimpleITK \cite{lowekamp2013} toolkit. 

All the compared methods used the same dataset composition and pre-processing operations. They were implemented in Python with the PyTorch-based medical image framework MONAI\footnote{More details about how to use those models can be found in MONAI core tutorials \url{https://monai.io/started.html##monaicore}.}.

\paragraph{Parameter initialization and learning}
At the FE stage, the number of filters in UNet was set to $(8, 16, 32, 64, 128)$ with kernel size equal to five and convolutional strides equal to $(2, 2, 2, 2)$ for layers from left to right. For nnUNet used in Section \ref{subsec: brain}, the kernel size was set to $(3, (1,1, 3), 3, 3)$ and the upsample kernel size was set to $(2,2,1)$ with strides $((1,1,1), 2, 2, 1)$. For nnFormer used in Section \ref{subsec: brain}, the crop size was set to $(128, 128, 128)$ with embedding dimension set to $96$ and the number of heads was set to $(3, 6, 12, 24)$. The number of extracted features was  $H=2$ for the PET-CT lymphoma dataset and  $H=4$ for the multi-MRI BraTS2021 dataset. 

To train our fusion framework, we proceeded in three steps. First, FE modules (i.e., UNet, nnUNet, or nnFormer) were pre-trained independently for each modality during 50 epochs. Then, the weights of the FE modules were fixed, and the parameters of the EM and MMEF modules were optimized. Finally, the whole framework was fine-tuned for a few epochs. The initial values of parameters $\alpha_i$ and $\gamma_i$ in the EM modules were set to 0.5 and 0.01, and the membership degrees $u_{ik}$ were initialized randomly by drawing uniform random numbers, and normalizing. We used, $I=10$ prototypes for the PET-CT lymphoma dataset, and $I=20$ prototypes for the more complex multi-MRI BraTS2021 dataset. These prototypes were randomly initialized from a normal distribution with zero mean and an identity covariance matrix. Details about the initialization of the EM module can be found in \cite{huang2022lymphoma}. The reliability coefficients $\beta^t_k$ in the MMEF module were initialized at 0.5. 

For both datasets, we used the Adam optimization algorithm with an early stopping strategy: training was stopped when there was no improvement in performance on the validation set during ten epochs. The initial learning rate was set to $0.01$. The batch size was set to 4. For all the compared methods, the model with the best performance on the validation set was saved as the final model for testing\footnote{The code is available at \url{https://github.com/iWeisskohl/Deep-evidential-fusion}.}. %\new{All methods were implemented in Python with the PyTorch-based medical image framework MONAI, and were trained and tested on a desktop with NVIDIA A100-SXM4 graphics card with 40 GB GPU memory.}
%The deep evidential fusion framework was implemented in Python with the PyTorch-based medical image framework MONAI and was trained and tested on a desktop with a 2.20GHz Intel(R) Xeon(R) CPU E5-2698 v4 and a Tesla V100-SXM2 graphics card with 32 GB GPU memory.

\paragraph{Evaluation criteria}
Although many authors have shown that segmentation performance can be improved by merging multimodal medical images into deep neural networks \cite{qu2023qnmf, andrade2023multi}, \new{the reliability of information sources and the quality of uncertainty quantification have rarely been investigated. Here, the former issue will be addressed by analyzing the reliability coefficients $\beta_k^t$ defined in Section \ref{subsubsec:MMEF}. To assess the quality of uncertainty quantification, we will use three metrics:} the \emph{Brier score} \cite{brier1950verification}, the \emph{negative log-likelihood} (NLL), and \emph{Expected Calibration Error} (ECE) \cite{guo2017calibration}. These metrics provide a robust evaluation framework for the uncertainty of the segmentation results, with smaller values indicating better performance. Their definitions are recalled below.

The Brier Score and NLL are defined, respectively, as 
\begin{equation*}
    \BS=\frac{1}{N} \sum_{n=1}^{N} (P_{n}-G_{n})^2,
%    \label{eq: bs}
\end{equation*}
and
\begin{equation*}
    \NLL=-\sum_{n=1}^{N}G_n \log P_n +(1-G_n)\log(1-P_{n}),
%    \label{eq: nll}
\end{equation*}
where $G_n$ is the ground truth of voxel $n$, $P_{n}$ is the predicted probability of voxel $n$, and $N$ is the number of voxels. 

The ECE measures the correspondence between predicted probabilities and ground truth. The output normalized plausibilities of the model are first discretized into equally spaced bins $E_b$, $b \in [1, B]$ ($B=10$ in this paper). The accuracy of bin $E_b$ is defined as
 \begin{equation*}
     \acc(E_b)=\frac{1}{\mid E_b\mid}\sum_{n \in E_b} \boldsymbol{1} (S_{n}=G_{n}),
 \end{equation*}
 where $S_n$ is the predicted class label for voxel $n$ and $ \boldsymbol{1}(\cdot)$ is the indicator function. The average confidence of bin $E_b$ is defined as
 \begin{equation*}
\conf(E_b)=\frac{1}{\mid E_b \mid}\sum_{n \in E_b}  P_{n}.
 \end{equation*}
The ECE is the weighted average of the difference in accuracy and confidence of the bins:
\begin{equation*}
\ECE= \sum_{b=1}^{B} \frac{\mid E_b \mid }{N}\mid \acc(E_b)-\conf(E_b)\mid.
%\label{eq:ece}
\end{equation*}
A model is perfectly calibrated when $\acc(E_b)=\conf(E_b)$ for all $b\in \{1,...,B\}$, in which case $\ECE=0$. 

Since our dataset has imbalanced foreground and background proportions, we only considered voxels belonging to the foreground or tumor region to calculate the above three indices. For the PET-CT lymphoma dataset, focusing only on the tumor region is not easy since the lymphomas are scattered throughout the whole body. Thus, we focused on the foreground region for this dataset. For the BraTS2021 dataset, we followed the suggestion from \cite{rousseau2021post} to focus on the tumor region for the reliability evaluation. For each patient in the test set, we defined a bounding box covering the foreground or tumor region and calculated the corresponding values in this bounding box. For all segmentation performance criteria, the reported results were obtained by calculating the criteria for each test 3D scan and then averaging over the patients.

In addition to evaluating segmentation reliability, we also measured segmentation accuracy using the \emph{Dice score}. In a segmentation task, the Dice score measures the volume of the overlapping region of the predicted object and the ground truth object  as 
\begin{equation*}
    \textsf{Dice}=\frac{2\, TP}{FP+2\, TP+FN}, 
\end{equation*}
where $TP$, $FP$, and $FN$ denote, respectively, the numbers of true positive, false positive, and false negative voxels.

\subsection{Segmentation results on the PET-CT lymphoma dataset}
\label{subsec: lym}
 
\paragraph{Segmentation uncertainty}
The results concerning uncertainty estimation are reported in Table \ref{tab:compar-lym}. Our model (MMEF-UNet) was compared to 
\begin{enumerate}
\item UNet with a softmax decision layer (the baseline); 
\item UNet with Monte-Carlo (MC) dropout \cite{gal2016dropout} and deep ensemble \cite{lakshminarayanan2017simple}, two popular techniques for improving the uncertainty quantification capabilities of probabilistic deep neural networks;
\item ENN-UNet, composed of UNnet as the FE module and the EM module in place of the softmax layer; this is the architecture studied in \cite{huang2022lymphoma};
\item RBF-UNet, an alternative model composed of UNnet and a radial-basis function (RBF) module in place of the softmax layer; as shown in \cite{huang2022lymphoma}, this model makes it possible to compute output belief functions that are similar to those computed by ENN-UNet.
\end{enumerate} 
We can remark that approaches 1 to 4 above implement pixel-level fusion, whereas our approach is based on decision-level fusion. \new{As for uncertainty quantification,  UNet, UNet-MC and UNet-Ensemble are probabilistic methods. UNet only computes point estimates of class probabilities without taking into account second-order uncertainty. UNet-MC applies dropout during both training and inference, sampling multiple forward passes to estimate uncertainty by averaging the predictions. UNet-Ensemble quantifies uncertainty by averaging the predictions obtained from multiple independently-trained models. In contrast, ENN-UNet and RBF-UNet are evidential methods: they both calculate belief functions to represent segmentation evidence and uncertainty under the DST framework.} For UNet-MC, the dropout rate was set to 0.2 and the number of samples was set to five; we averaged the five output probabilities at each voxel as the final output of the model. For UNet-ensembles, the number of samples was set to five; the five output probabilities were then averaged at each voxel as the final output of the model. \new{The settings of ENN-UNet and RBF-ENN are the same as those reported in \cite{huang2022lymphoma}.}

\new{From Table \ref{tab:compar-lym}, we can see that Monte-Carlo dropout and deep ensembles do not significantly improve the segmentation reliability as compared to the baseline UNet model, as shown, e.g., by the higher NLL values. In contrast, the addition of the EM module to the FE module, as implemented in ENN-UNet, brings a significant improvement, particularly according to NLL; the RBF-UNet model yields similar results. The decision-fusion framework MMEF-UNet brings an additional improvement according to all three criteria (ECE, Brier score, NLL) and outperforms the other models: specifically, we observe decreases of 1.1\%, 0.9\%, and 13\% in ECE, Brier score, and NLL, respectively, as compared to UNet.} We can conclude that, compared to the baseline model, both the EM and MMEF modules contribute to a higher segmentation reliability.  

\begin{table}
\caption{Means and standard errors of segmentation quality and reliability measures for MMEF-UNet and the referenced uncertainty quantification methods on the lymphoma dataset. The best results are in bold and the second best are underlined.}
\begin{center}
\scalebox{0.9}{
    \begin{tabular}{ccccc}
    \hline
     Model   &\ECE  $\downarrow$  &Brier score $\downarrow$ &\NLL $\downarrow$ &Dice score $\uparrow$\\
\hline
%UNet    &0.056$\pm$0.008  &0.065$\pm$0.009  &0.310$\pm$0.196 &0.770$\pm$0.063\\
UNet    &0.056$\pm$3.6$\times10^{-3}$ &0.065$\pm$3.9$\times10^{-3}$  &0.310$\pm$8.8$\times10^{-2}$  &{0.770}$\pm$3.2$\times10^{-2}$ \\
%UNet-MC   &0.053$\pm$0.009  &0.062$\pm$0.011  &0.400$\pm$0.195 &0.801$\pm$0.023\\
UNet-MC   &0.053$\pm$4.6$\times10^{-3}$   &{0.062}$\pm$4.9$\times10^{-3}$   &0.400$\pm$8.7$\times10^{-2}$ &0.801$\pm$1.1$\times10^{-2}$\\
%UNet-Ensemble  &0.063$\pm$0.017  &0.064$\pm$0.009  &0.343$\pm$0.162 &0.802$\pm$0.015 \\
UNet-Ensemble  &0.063$\pm$7.6$\times10^{-2}$  &0.064$\pm$4.0$\times10^{-3}$  &0.343$\pm$7.2$\times10^{-2}$ &0.802$\pm$6.7$\times10^{-3}$ \\
%ENN-UNet   &0.050$\pm$0.008  &0.062$\pm$0.009  &0.191$\pm$0.031 &0.805$\pm$0.016 \\
ENN-UNet  &\underline{0.050}$\pm$3.5$\times10^{-3}$  &{0.062}$\pm$3.9$\times10^{-3}$  &\underline{0.191}$\pm$1.4$\times10^{-2}$ &\underline{0.805}$\pm$7.1$\times10^{-3}$  \\
RBF-UNet  &0.051$\pm$3.3$\times10^{-3}$  &\underline{0.061}$\pm$0.9$\times10^{-3}$  &{0.193}$\pm$1.3$\times10^{-2}$ &{0.802}$\pm$6.9$\times10^{-3}$  \\
%nnUNet &0.036$\pm$0.014 & 0.057$\pm$0.013 &0.208$\pm$0.137 &0.783$\pm$0.070 \\
%MMEF-UNet (ours)  &0.045$\pm$0.003  &0.056$\pm$0.006  &0.180$\pm$0.028 &0.811$\pm$0.078 \\
MMEF-UNet (ours) &\textbf{0.045}$\pm$1.3$\times10^{-3}$    &\textbf{0.056}$\pm$2.7$\times10^{-3}$ &\textbf{0.180}$\pm$1.3$\times10^{-2}$&\textbf{0.811}$\pm$$3.4\times10^{-2}$\\
%MMEF-nnUNet (ours)  &\textbf{0.030}$\pm$0.010  &\textbf{0.051}$\pm$0.009  &\textbf{0.159}$\pm$0.048 &\textbf{0.819}$\pm$0.076 \\
\hline
\end{tabular}
}
\end{center}
\label{tab:compar-lym}
\end{table}

\begin{figure}
    \centering
    \includegraphics[width=\textwidth]{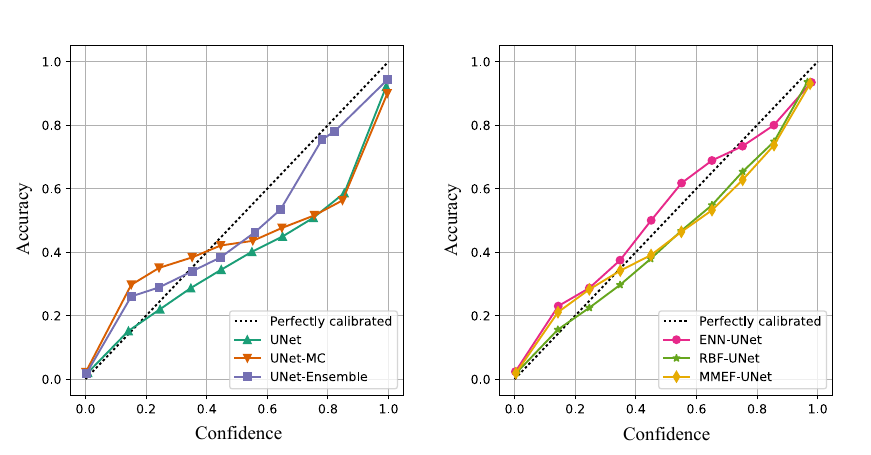}
    \caption{Calibration plots for probabilistic (left) and evidential (right)  deep segmentation models.}
    \label{fig:calibration-plots-lym}
\end{figure}

These findings are, to some extent, confirmed by Figure \ref{fig:calibration-plots-lym}, which shows the calibration plots (also known as reliability diagrams) for the compared methods on the lymphoma dataset. \new{Calibration plots are graphical representations showing how well the probabilistic predictions of a segmentation model are calibrated, i.e., how well confidence matches accuracy. In the left graph of Figure \ref{fig:calibration-plots-lym}, we can see that the curve corresponding to UNet-Ensemble is closer to the diagonal than those of UNet and UNet-MC, which indicates better calibration.} Looking at the right graph in Figure \ref{fig:calibration-plots-lym}, we can see that the three DST-based models, ENN-UNet, RBF-UNet, and MMEF-UNet, have better calibration performance than the probabilistic ones, as shown by their calibration curves closer to the diagonal. Among them, MMEF-UNet shows the best calibration performance as ENN-UNet is slightly overconfident, while RBF-UNet is slightly underconfident. 

\begin{figure}
    \centering
    \includegraphics[width=\textwidth]{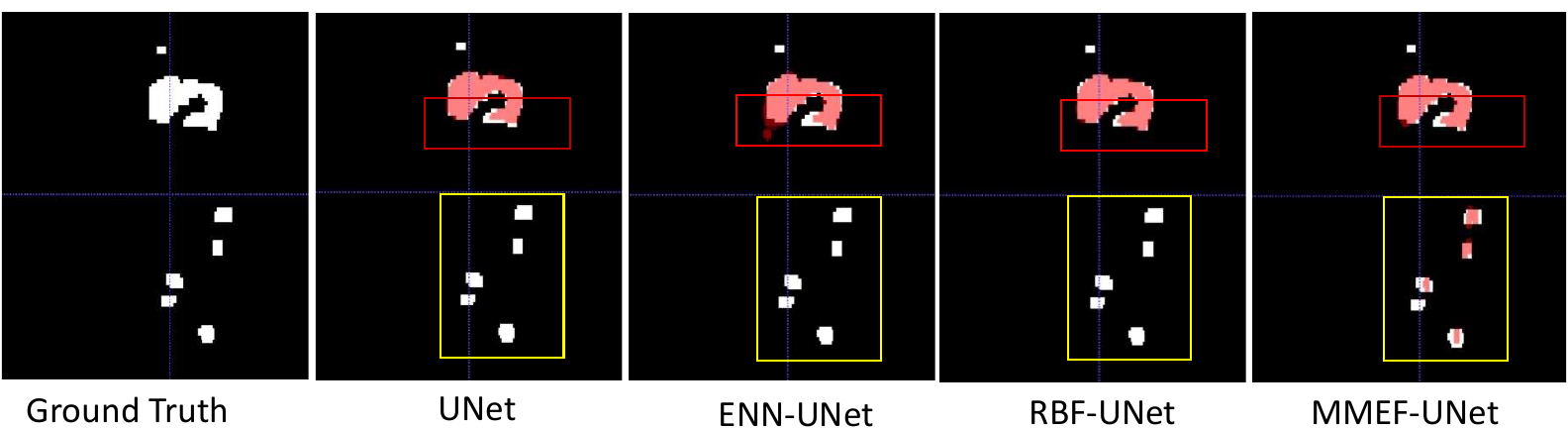}
    \caption{Examples of visualized segmentation results: from left to right,  ground truth, and segmentation results obtained by UNet, ENN-UNet, RBF-UNet, and MMEF-UNet. The white and red regions represent, respectively, the ground truth and the segmentation result. Red and yellow boxes highlight the main differences in segmenting large and small isolated tumors.}
    \label{fig:accuracy_imshow}
\end{figure}

%%%%%%%%%%%%%%%%%%%%%%%%%%%%%%%%
\paragraph{Segmentation accuracy}
%\label{subsubsec: acc-lym}
The segmentation accuracy was measured by the Dice score, as shown in Table~\ref{tab:compar-lym}. Compared with the baseline model UNet, our proposal MMEF-UNet significantly increases segmentation performance, as shown by a 4.1\% increase in the Dice score. Compared with the two DST-based deep evidential segmentation methods, MMEF-UNet has a higher Dice score (although the difference with ENN-UNet is not statistically significant). Figure \ref{fig:accuracy_imshow} shows an example of visualized segmentation results obtained by UNet, ENN-UNet, RBF-UNet, and MMEF-UNet. We can see that UNet and RBF-UNet are more conservative (they correctly detect only a subset of the tumor voxels), while ENN-UNet is more radical (some of the voxels that do not belong to tumors are predicted as tumors). In contrast, the tumor regions predicted by MMEF-UNet better overlap the ground-truth tumor region, especially for the isolated lymphomas, which is also reflected by the promising Dice score value. These conclusions are consistent with the calibration trends displayed in Figure \ref{fig:calibration-plots-lym}. %\textbf{The best segmentation accuracy was achieved by nnUNet-MMEF, with the highest Dice score of 0.819 and a 1.4\% increase in Dice score compared with the second-best method, ENN-UNet. }

\paragraph{Analysis of reliability coefficients} 

%\begin{figure}
%    \centering
%    \includegraphics[width=0.8\textwidth]{figure/Coefficients_lymphoma.pdf}
%    \caption{\new{Estimated reliability coefficient $\beta_k$ (means and standard errors) after training for the background and lymphoma classes and the two modalities. Higher values correspond to greater contribution to the segmentation.}}
%    \label{fig: beta-lym}
%\end{figure}

\begin{table}
 \caption{Estimated reliability coefficient $\beta^t_k$ (means and standard errors) after training for the background and lymphoma classes and the two modalities. Higher values correspond to greater contribution to the segmentation.}
  \begin{center}
  \label{tab: beta-lym}
  \begin{tabular}{lccc}
  \hline
  \multicolumn{1}{l}{$\beta_k^t$}& \multicolumn{1}{l}{background}  & \multicolumn{1}{l}{lymphomas} \\
  \hline
PET &0.999$\pm$8.9$\times10^{-3}$   &0.996$\pm$4.5$\times10^{-3}$  \\
CT  &0.863$\pm$1.8$\times10^{-2}$   &0.975$\pm$8.9$\times10^{-3}$  \\
\hline
\end{tabular}
\end{center}
\end{table}

Table \ref{tab: beta-lym} reports the learned reliability coefficients. We can see that they are higher for the PET modality. This is consistent with domain knowledge, as mentioned in Section \ref{subsec: setting}: PET images provide functional information about tumor activity and make it possible to \new{identify active tumor sites}, whereas CT images essentially provide detailed anatomical information (e.g., size, shape, and location) about lymph nodes and surrounding tissues and are used as a complement to PET images. This is also confirmed by the results presented in Table \ref{tab:modalities}, showing that the performance of UNet and ENN-UNet with either CT alone or PET alone, the latter configuration yielding better results.
\begin{table}
\caption{Segmentation quality and reliability of UNet and ENN-UNet applied to the lymphoma dataset with a single modality (CT or PET).}
\begin{center}
\scalebox{0.92}{
\begin{tabular}{cccccccc}
\hline
Model &\ECE $\downarrow$ &Brier score $\downarrow$ &\NLL $\downarrow$ &Dice score $\uparrow$  \\
 \cline{3-5} 
\hline
UNet (CT)& 0.133$\pm$4.9$\times10^{-3}$ &0.157$\pm$9.8$\times10^{-3}$ &0.571$\pm$3.6$\times10^{-2}$&0.544$\pm$2.8$\times10^{-2}$\\
UNet (PET) &0.060$\pm$4.0$\times10^{-3}$ &0.068$\pm$4.0$\times10^{-3}$ &0.348$\pm$8.2$\times10^{-2}$ &0.764$\pm$2.9$\times10^{-2}$ \\
\hline
ENN-UNet (CT)   &0.131$\pm$8.5$\times10^{-3}$ &0.156$\pm$1.0$\times10^{-2}$&0.521$\pm$3.2$\times10^{-2}$ &0.543$\pm$2.7$\times10^{-2}$  \\
ENN-UNet (PET)  &{0.050}$\pm$4.9$\times10^{-3}$ &0.064$\pm$5.4$\times10^{-3}$ &0.195$\pm$2.2$\times10^{-2}$ &{0.781}$\pm$3.5$\times10^{-2}$  \\
\hline
\end{tabular}
}
\end{center}
\label{tab:modalities}
\end{table}

%%%%%%%%%%%%%%%%%%%%%%%%%%%%%%
\subsection{Segmentation results on the multi-MRI BraTS2021 dataset}
\label{subsec: brain}
\paragraph{Segmentation uncertainty} 

For the BraTS2021 dataset, we tested the segmentation performance of our fusion framework with UNet as well as two alternative FE modules:  nnUNet \new{and nnFormer}. The nnUNet model was reported to have the best performance in the BraTS2021 challenge \cite{luu2021} and \new{nnFormer is now one of the state-of-the-art brain tumor segmentation models}. The complete frameworks with nnUNet \new{and nnFormer} as a feature extractor are referred to, respectively,  as MMEF-nnUNet \new{and MMEF-nnFormer}.
%\textbf{For nnUNet, the kernel size was set as $(3, (1,1, 3), 3, 3)$ and the upsample kernel size was set as $(2,2,1)$ with strides $((1,1,1), 2, 2, 1)$.} 
We compared our results with three baseline models: \new{UNet, nnUNet, and nnFormer}, and three Monte Carlo-based uncertainty segmentation models: \new{UNet-MC, nnUNet-MC, and nnFormer-MC}. Since the results obtained in Section \ref{subsec: lym}, as well as those reported in \cite{huang2022lymphoma} have shown that ENN-UNet and RBF-UNet yield similar results, here we only compared the performance of \new{the ENN-based models, i.e., ENN-UNet, ENN-nnUNet and ENN-nnFormer}. Moreover, we did not test the performance of deep ensemble models because applying them to larger-scale datasets exceeds our computation resources.

\begin{table}
\caption{Reliability measures (means and standard errors) for MMEF-UNet and the reference methods based on UNet on the BraTS2021 dataset. The best results are in bold and the second bests are underlined.}
\begin{center}
%\scalebox{0.7}{   
%{\footnotesize
\begin{tabular}{cccc}
\hline
Model &\ECE $\downarrow$ &Brier score $\downarrow$ &\NLL $\downarrow$  \\
\hline
UNet   &0.071$\pm$1.8$\times10^{-3}$  &0.141$\pm$1.8$\times10^{-3}$ &2.475$\pm$2.2$\times10^{-3}$ \\
UNet-MC   &0.067$\pm$1.3$\times10^{-3}$  &0.135$\pm$4.5$\times10^{-3}$ &2.264$\pm$7.3$\times10^{-2}$  \\
ENN-UNet  &\underline{0.065}$\pm$1.3$\times10^{-3}$  &\underline{0.130}$\pm$4.5$\times10^{-3}$  & \underline{2.250}$\pm$3.6$\times10^{-2}$  \\
MMEF-UNet (ours) &\textbf{0.060}$\pm$1.3$\times10^{-3}$  &\textbf{0.115}$\pm$2.2$\times10^{-3}$  &\textbf{2.189}$\pm$4.1$\times10^{-2}$    \\
\hline
\end{tabular}
%}
\end{center}
\label{tab: reliability-brats_UNet}
\end{table}

\begin{table}
        \caption{Reliability measures (means and standard errors) for MMEF-nnUNet and the reference methods based on nnUNet on the BraTS2021 dataset. The best results are in bold. The best results are in bold, and the second-best results are underlined.}
        \begin{center}
%\scalebox{0.7}{   
%{\footnotesize
    \begin{tabular}{cccc}
    \hline
     Model &\ECE $\downarrow$ &Brier score $\downarrow$ &\NLL $\downarrow$  \\
\hline
nnUNet  & \underline{0.053}$\pm$2.2$\times10^{-3}$  & 0.109$\pm$4.5$\times10^{-3}$ &1.823$\pm$7.3$\times10^{-2}$    \\
nnUNet-MC & \textbf{0.051}$\pm$1.8$\times10^{-3}$ & \underline{0.107}$\pm$4.5$\times10^{-3}$ &1.810$\pm$5.8$\times10^{-2}$  \\
ENN-nnUNet &\underline{0.053}$\pm$1.8$\times10^{-3}$  &0.109$\pm$4.9$\times10^{-3}$ &\underline{1.804}$\pm$8.2$\times10^{-2}$   \\
MMEF-nnUNet (ours) &\textbf{0.051}$\pm$1.3$\times10^{-3}$  & \textbf{0.102}$\pm$2.7$\times10^{-3}$  & \textbf{1.748}$\pm$5.9$\times10^{-2}$ \\
\hline
\end{tabular}
%}
\end{center}
\label{tab: reliability-brats_nnUNet}
\end{table}

\begin{table}
\caption{\new{Reliability measures (means and standard errors) for MMEF-nnFormer and the reference methods based on nnUNet on the BraTS2021 dataset. The best results are in bold and the second-best are underlined.}}
\begin{center}
%\scalebox{0.7}{   
%{\footnotesize
\new{
\begin{tabular}{cccc}
\hline
Model &\ECE $\downarrow$ &Brier score $\downarrow$ &\NLL $\downarrow$  \\
\hline
nnFormer  & 0.055$\pm$1.6$\times10^{-3}$  & 0.111$\pm$3.2$\times10^{-3}$ &1.917$\pm$5.5$\times10^{-2}$    \\
%0.055507895	0.11101579	1.917174766
%0.001602558	0.003205116	0.0553504
nnFormer-MC & \underline{0.053}$\pm$1.8$\times10^{-3}$ & \underline{0.107}$\pm$3.6$\times10^{-3}$ &\textbf{1.756}$\pm$6.1$\times10^{-2}$  \\
%0.053434848	0.107245623	1.756268557
%0.001817349	0.003603015	0.060689054
ENN-nnFormer &0.055$\pm$1.4$\times10^{-3}$  &0.110$\pm$3.6$\times10^{-3}$ &1.907$\pm$7.0$\times10^{-2}$   \\
%0.055403238	0.109606475	1.906637585
%0.001454606	0.003560025	0.070191305
MMEF-nnFormer (ours) &\textbf{0.052}$\pm$0.6$\times10^{-3}$  & \textbf{0.103}$\pm$1.2$\times10^{-3}$  & \underline{1.787}$\pm$2.2$\times10^{-2}$ \\
%0.05173132	0.10346264	1.786736502
%0.000643446	0.001286893	0.022223854
\hline
\end{tabular}
}
\end{center}
\label{tab: reliability-brats_nnFormer}
\end{table}
As with the lymphoma dataset, we used the ECE, Brier score, and NLL metrics to assess segmentation uncertainty. The results with UNet, nnUNet \new{and nnFormer} in the FE module are presented, respectively, in Tables \ref{tab: reliability-brats_UNet}, \ref{tab: reliability-brats_nnUNet} \new{and \ref{tab: reliability-brats_nnFormer}}. We can see that our fusion model consistently outperforms the baseline models with all three FE models and across all uncertainty evaluation metrics, although the \new{differences} are more significant when UNet is used as a feature extractor. 
Indeed, the fusion mechanism can be expected to have a smaller impact when information sources are more informative. \new{Overall, MMEF-nnUNet achieves the highest segmentation reliability with the lowest ECE, Brier score, and NLL values, and MMEF-nnFormer yields the second-best results.}

\paragraph{Segmentation accuracy}
%\label{subsubsec: acc-brats}
Segmentation accuracy was evaluated by the Dice score for the three overlapping regions, ET, TC, and WT, as well as by the mean Dice score. The results with UNet, nnUNet \new{and nnFormer} as feature extractors are reported, respectively, in Tables~\ref{tab: dice-brats_UNet}, \ref{tab: dice-brats_nnUNet} \new{and \ref{tab: dice-brats_nnFormer}. Again, we can see that our fusion strategy improves segmentation accuracy for all three FE models. Overall, the highest segmentation accuracy was achieved by MMEF-nnFormer, with an increase of 1.5 \% in the mean Dice score compared with the second-best method, ENN-nnFormer. } 

\new{We also report the Dice score for the segmentation of the three original tumor regions: ED, ET, and NRC/NET in Table \ref{tab: dice-brats_nnFormer_subclass}. As we can see, the baseline nnFormer shows good performance for segmenting ED and ET, while it does not perform as well for segmenting NRC/NET. Indeed, the lack of clear contrast, the similar signal intensities to normal brain tissue, the infiltrative growth patterns, and the need for multi-modal data make the segmentation of NRC/NET inherently more challenging compared to ED and ET. When the MMEF-nnFormer approach was applied, the Dice scores for the ED, ET, and NRC/NET improved by 0.6\%, 1.6\%, and 6.5\%, respectively. The substantial improvement in  NRC/NET segmentation is particularly encouraging, as it demonstrates the effectiveness of the proposed fusion method for delineating fuzzy tumor boundaries and solving challenging segmentation tasks.
}

\begin{table}
\caption{Dice score (means and standard errors) for MMEF-UNet and the reference methods based on UNet on the BraTS2021 dataset. The best results are in bold and the second bests are underlined.}
\begin{center}
\scalebox{0.87}{   
%{\footnotesize
\begin{tabular}{ccccc}
\hline
Model  &ET&TC&WT&Mean\\
\hline

%UNet   & 0.807$\pm$0.021 &0.825$\pm$0.019& 0.881$\pm$0.015&0.837$\pm$0.016\\
%UNet-MC   &0.812$\pm$0.029 &0.832$\pm$0.024& 0.886$\pm$0.014& 0.843$\pm$0.020\\
%ENN-UNet  &0.810$\pm$0.029 &0.842$\pm$0.024& 0.896$\pm$0.012& 0.849$\pm$0.021\\
%MMEF-UNet (ours) & 0.833$\pm$0.027 &0.854$\pm$0.016 &0.907$\pm$0.011 &0.864$\pm$0.013 \\

UNet   & 0.807$\pm$9.4$\times10^{-3}$ &0.825$\pm$8.5$\times10^{-3}$& 0.881$\pm$6.7$\times10^{-3}$ &0.837$\pm$7.2$\times10^{-3}$\\
UNet-MC   &0.812$\pm$1.3$\times10^{-2}$ &0.832$\pm$1.1$\times10^{-2}$ & 0.886$\pm$6.3$\times10^{-3}$ & 0.843$\pm$8.9$\times10^{-3}$\\
ENN-UNet  &\underline{0.810}$\pm$1.3$\times10^{-2}$ &\underline{0.842}$\pm$1.1$\times10^{-2}$& \underline{0.896}$\pm$5.4$\times10^{-3}$& \underline{0.849}$\pm$9.4$\times10^{-3}$\\
MMEF-UNet (ours) & \textbf{0.833}$\pm$1.2$\times10^{-2}$ &\textbf{0.854}$\pm$7.2$\times10^{-3}$ &\textbf{0.907}$\pm$4.9$\times10^{-3}$  &\textbf{0.864}$\pm$5.8$\times10^{-3}$  \\
\hline
\end{tabular}
}
\end{center}
\label{tab: dice-brats_UNet}
\end{table}

\begin{table}
        \caption{Dice score (means and standard errors) for MMEF-nnUNet and the reference methods based on nnUNet on the BraTS2021 dataset. The best results are in bold and the second bests are underlined.}
        \begin{center}
\scalebox{0.87}{   
%{\footnotesize
    \begin{tabular}{ccccc}
    \hline
     Model  &ET&TC&WT&Mean\\
\hline
nnUNet  & 0.791$\pm$4.9$\times10^{-3}$ & 0.850$\pm$5.8$\times10^{-3}$&	0.912$\pm$3.5$\times10^{-3}$ &0.851$\pm$4.4$\times10^{-3}$ \\
nnUNet-MC &0.802$\pm$4.4$\times10^{-3}$&	0.860$\pm$4.9$\times10^{-3}$&	\underline{0.916}$\pm$4.4$\times10^{-3}$ &0.859$\pm$4.9$\times10^{-3}$\\
ENN-nnUNet &\underline{0.807}$\pm$9.8$\times10^{-3}$  & \underline{0.869}$\pm$1.9$\times10^{-2}$ & 0.915$\pm$5.4$\times10^{-3}$ &\underline{0.863}$\pm$9.8$\times10^{-3}$\\
MMEF-nnUNet (ours)  &\textbf{0.832}$\pm$9.8$\times10^{-3}$ & \textbf{0.873}$\pm$2.6$\times10^{-3}$ & \textbf{0.918}$\pm$1.3$\times10^{-3}$ & \textbf{0.875}$\pm$4.4$\times10^{-3}$\\
\hline
\end{tabular}
}
\end{center}
\label{tab: dice-brats_nnUNet}
\end{table}

\begin{table}
\caption{\new{Dice score (means and standard errors) for MMEF-nnFormer and the reference methods based on nnFormer on the BraTS2021 dataset. The best results are in bold and the second bests are underlined.}}
        \begin{center}
\scalebox{0.85}{   
\new{
    \begin{tabular}{ccccc}
    \hline
     Model  &ET&TC&WT&Mean\\
\hline
nnFormer  & \underline{0.839}$\pm$3.8$\times10^{-3}$ & 0.878$\pm$2.9$\times10^{-3}$&	\textbf{0.915}$\pm$2.4$\times10^{-3}$ &0.877$\pm$1.7$\times10^{-3}$ \\ 
%0.83924	0.87822	0.91474	0.8774
%0.003765979 0.002984527	0.002488092	0.001744356
nnFormer-MC &0.837$\pm$3.7$\times10^{-3}$&	0.877$\pm$4.5$\times10^{-3}$&	\underline{0.914}$\pm$2.9$\times10^{-3}$ &0.876$\pm$2.3$\times10^{-3}$\\ 
%0.83686	0.87746	0.91386	0.87606
%0.003695754	0.00456537	0.002907336	0.002302882%
ENN-nnFormer &0.836$\pm$9.8$\times10^{-3}$  & \underline{0.882}$\pm$5.6$\times10^{-2}$ & \underline{0.914}$\pm$5.2$\times10^{-3}$ &\underline{0.878}$\pm$3.2$\times10^{-3}$\\
%0.83642	0.88218	0.91414	0.87758
%0.009847182	0.005620676	0.00518681	0.003269438
MMEF-nnFormer (ours)  &\textbf{0.854}$\pm$7.5$\times10^{-3}$ & \textbf{0.911}$\pm$5.4$\times10^{-3}$ & \underline{0.914}$\pm$2.3$\times10^{-3}$ & \textbf{0.893}$\pm$4.8$\times10^{-3}$\\
%0.8548	0.91146	0.9137	0.89332
%0.0075004	0.005456244	0.002354358	0.004847618
\hline
\end{tabular}
}
}
\end{center}
\label{tab: dice-brats_nnFormer}
\end{table}

\begin{table}
\caption{\new{Dice score (means and standard errors) for MMEF-nnFormer and nnFormer on the BraTS2021 dataset in segmenting detailed tumor class.}}
\begin{center}
\scalebox{0.85}{   
\new{
\begin{tabular}{ccccc}
\hline
Model  &ED&ET&NRC/NET&Mean\\
\hline
nnFormer  & 0.817$\pm$5.0$\times10^{-3}$ & 0.839$\pm$3.8$\times10^{-3}$&	0.740$\pm$7.2$\times10^{-3}$ &0.799$\pm$2.5$\times10^{-3}$ \\ 
%	0.81686	0.83924	0.73982 0.79864
%	0.005035335	0.003765979	0.007210576 0.002474973
MMEF-nnFormer (ours)  &\textbf{0.823}$\pm$3.3$\times10^{-3}$ & \textbf{0.855}$\pm$7.5$\times10^{-3}$ & \textbf{0.805}$\pm$7.3$\times10^{-3}$ & \textbf{0.828}$\pm$4.7$\times10^{-3}$\\
%	0.8227	0.8548	0.80516 0.827553333
%	0.003282073	0.0075004	0.007332162 0.004683512
\hline
\end{tabular}
}
}
\end{center}
\label{tab: dice-brats_nnFormer_subclass}
\end{table}

\begin{figure}
    \centering
    \includegraphics[width=\textwidth]{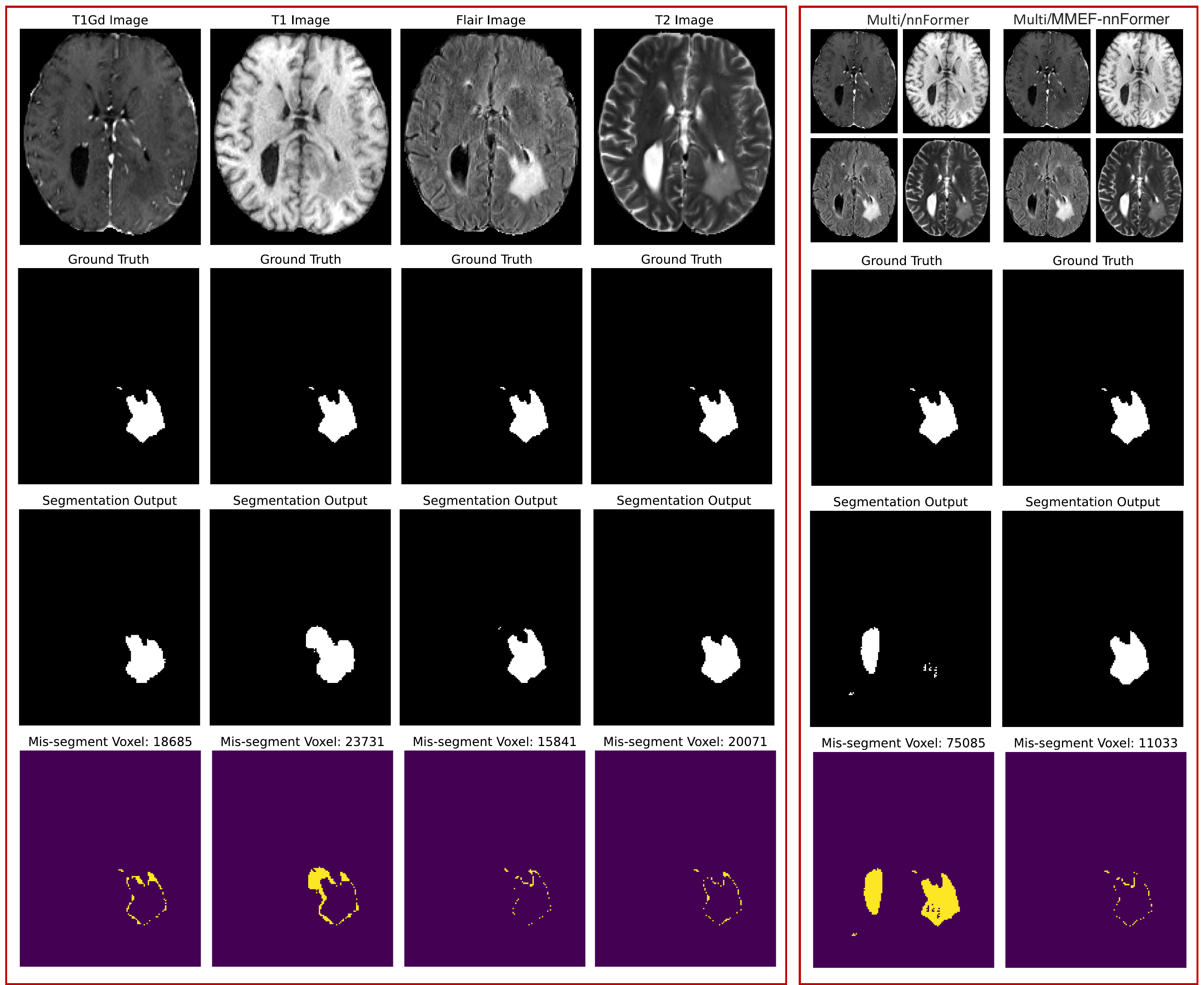}
    \caption{\new{Examples of an easy tumor segmentation case. The first and second rows display the input modalities and the tumor ground truth, respectively. The third and last rows present the segmentation output and the mis-segmented voxels (highlighted in yellow). The left red block shows results from single-modality input using nnFormer, while the right red block compares results from multimodal input using nnFormer (left column) and MMEF-nnFormer (right column).}}
    \label{fig: easy-example}
\end{figure}

\begin{figure}
    \centering
    \includegraphics[width=\textwidth]{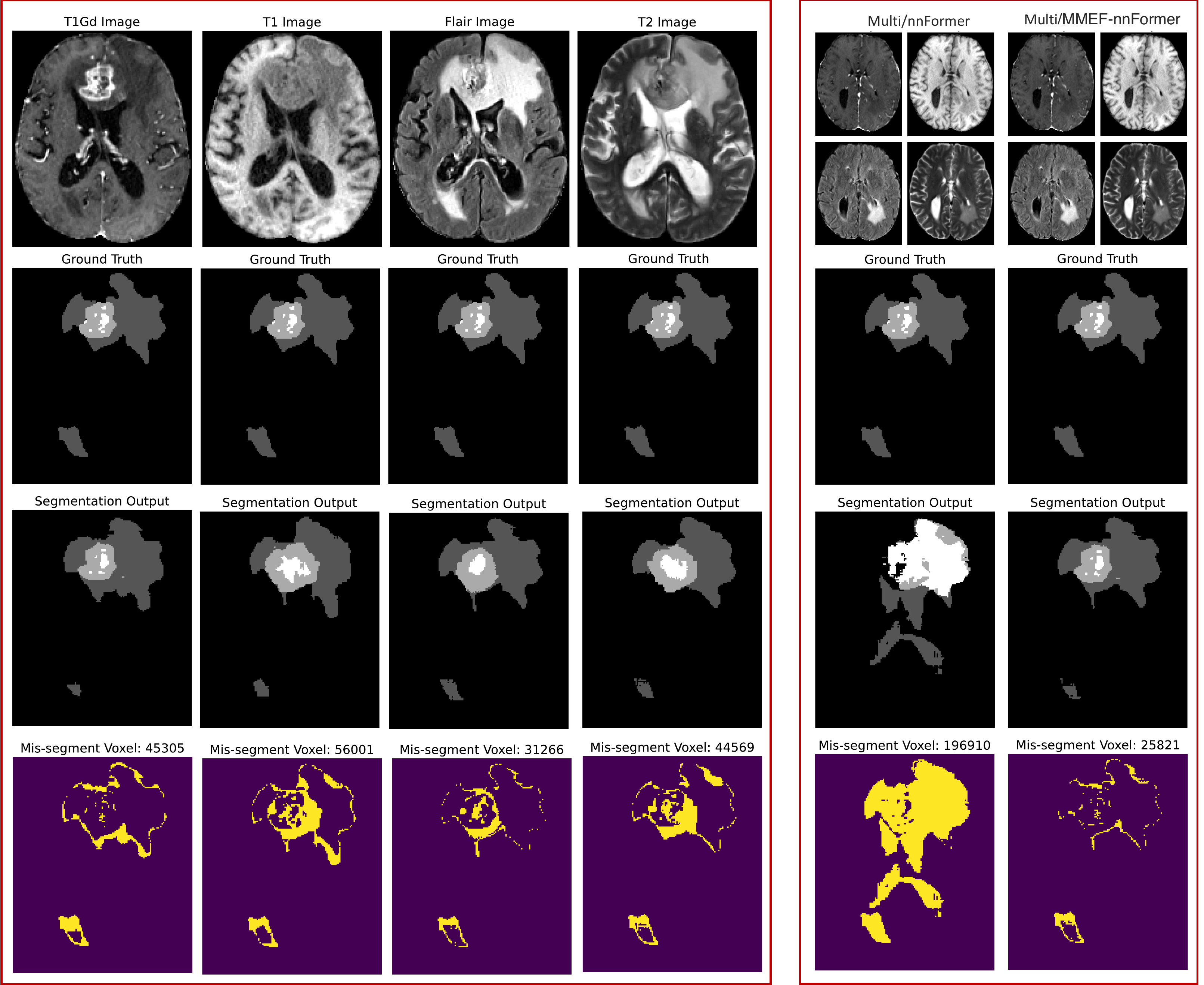}
    \caption{\new{Examples of a challenging tumor segmentation case. The first and second rows display the input modalities and the tumor ground truth, respectively. The third and last rows present the segmentation output and the mis-segmented voxels (highlighted in yellow). The left red block shows results from single-modality input using nnFormer, while the right red block compares results from multimodal input using nnFormer (left column) and MMEF-nnFormer (right column).} }
    \label{fig: hard-example}
\end{figure}

\new{Figures \ref{fig: easy-example} and \ref{fig: hard-example} show two segmentation cases when using nnFormer as the feature extractor. Figure \ref{fig: easy-example} shows an easy segmentation case where only one tumor type is present. Both the Flair and T1Gd images exhibit good segmentation performance with only a few mislabeled voxels. 
It is surprising to see that concatenating multimodal medical images as the input for nnFormer resulted in worse outcomes, with the most mislabeled voxels. This might be due to the hard fusion strategy of nnFormer, i.e., image concatenation, which cannot mitigate the impact of noisy information. Consequently, the fused results are sometimes not as good as those from single-modality inputs. The proposed MMEF-nnFormer approach achieves the best performance, with fewer mislabeled voxels compared to other methods.  Figure \ref{fig: hard-example} illustrates a challenging segmentation scenario involving a tumor with ED, ET, and NRC/NET components. We can remark that the FLAIR image alone provides sufficient information to accurately segment  ED, which is consistent with domain knowledge. Overall, the MMEF-nnFormer model yields the best results in this case.  This example illustrates the ability of our method to improve segmentation accuracy by appropriately weighting and combining information from different modalities.
}

\paragraph{Analysis of reliability coefficients} 
%\label{subsubsec:reliability}

\new{We first recall some clinical domain knowledge of MRI images in segmenting brain tumors: 
\begin{enumerate}
\item T1Gd images are particularly useful for delineating tumor boundaries by making tumor regions hyperintense (bright);
\item FLAIR images help delineate tumor boundaries, assess tumor infiltration into surrounding brain tissue, and are particularly sensitive to peritumoral edema, which appears hyperintense (bright) on FLAIR sequences;
\item T2 images help delineate tumor extent, identify peritumoral edema, and assess the relationship between the tumor and surrounding brain structures;
\item Tumors typically appear hypointense (dark) on T1 images, while the contrast between the tumor and surrounding normal brain tissue may not always be sufficient for accurate segmentation.
\end{enumerate}
}

\begin{figure}
    \centering
    \includegraphics[width=\textwidth]{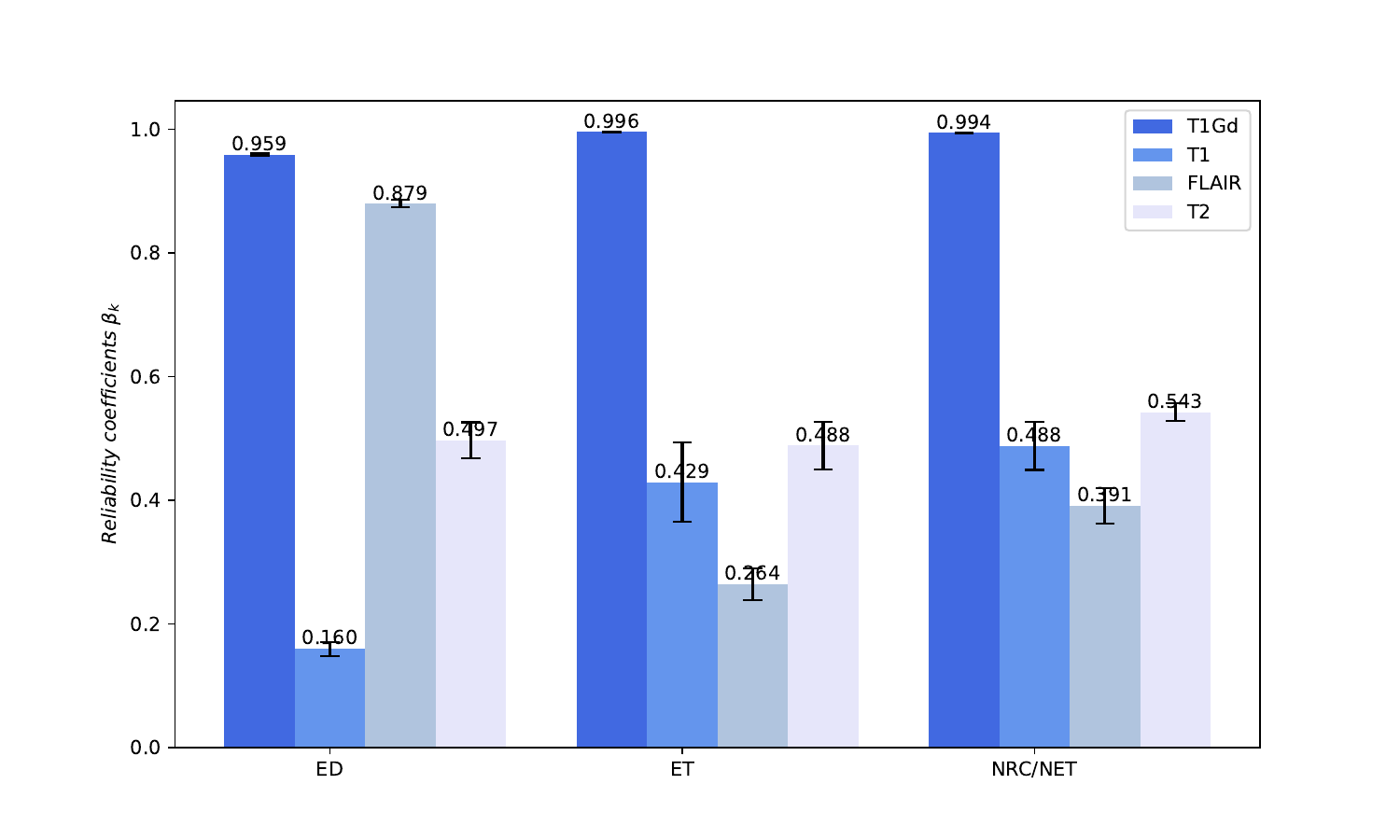}
    \caption{\new{Estimated reliability coefficients $\beta_k$ (means and standard errors) after training of MMEF-nnFormer for classes ED, ET, and NRC/NET in the four modalities. Higher values correspond to greater contribution to the segmentation.}}
    \label{fig: beta_nnformer}
\end{figure}

\new{Figure \ref{fig: beta_nnformer}} shows the learned reliability coefficients $\beta_k^t$ estimated by \new{MMEF-nnFormer}, for the four modalities and the three tumor classes. It can be seen that the evidence from the T1Gd modality is reliable when the true class is ED, ET, or NRC/NET, with all the reliability values greater than 0.9. \new{In contrast, the evidence from the FLAIR modality is more reliable for the ED class with a high-reliability coefficient of 0.879 against, respectively, 0.26 and 0.39 for ET and NRC/NET. The evidence from the T2 modality shows similar reliability in segmenting the three classes with a reliability coefficient of around 0.5. The evidence from the T1 modality is the least reliable one, compared with the other three MRI modalities. These results are consistent with domain knowledge about these modalities as reported in \cite{baid2021rsna} and recalled at the beginning of this section, i.e., T1Gd images are useful for delineating tumor boundaries, FLAIR images are sensitive to ED, and T1 images are not sufficient for accurate tumor segmentation. This transparency and explainability of the decision-making process can be expected to enhance end-users' trust and can be seen as significant advantages of the proposed multimodal evidence fusion approach, as opposed to the ``black box'' nature of conventional deep learning segmentation models.}

\new{
\subsection{Discussion}
\label{subsec:discussion}
In the following, we provide some discussion about the generalizability, computational complexity, and limitations of our approach.

\paragraph{Generalizability} 

The main advantage of our framework is its ability to model and learn the reliability of each image modality, which can be crucial when dealing with diverse, potentially noisy, or low-quality data. While multimodal medical image segmentation tasks are the focus of this paper, the proposed deep evidential fusion framework can be applied to a broader range of challenging medical tasks involving heterogeneous data sources. For instance, in medical tasks such as diagnosing dementia or Alzheimer's disease, various heterogeneous medical data are available \cite{bycroft2018uk}. These data can include lower-quality brain MRI images due to brain degeneration, textual data on disease history and progression, time-series data on blood-brain-barrier integrity, cerebrovascular information, and other relevant physiological measures. Traditional models struggle to effectively address this heterogeneous data within a single neural network \cite{dai2019research}, and recent work also proposed to address data heterogeneity with model ensembles and hard decision fusion \cite{tanveer2024ensemble}. Our deep evidential fusion framework could be well-suited to analyze such heterogeneous medical tasks. By learning the reliability coefficients for each of the modalities, our model can effectively combine the evidence from heterogeneous sources to reach a more informed and explainable diagnostic decision. 

Beyond medical image processing, our approach could be applied to multimodal data fusion in other domains, such as reviewed in \cite{lahat15} and \cite{bokade21}. As examples of potential application domains where heterogeneous data need to be processed to make decisions, we can mention remote sensing and earth observations, in which light detection and ranging (LiDAR), synthetic aperture radar (SAR), and hyperspectral images need to be combined for, e.g., improved classification of objects. As noted in \cite{lahat15}, SAR and LiDAR use different electromagnetic frequencies and thus interact differently with materials and surfaces. It would thus be beneficial to apply different discounting (reliability) coefficients to these sensor data depending on the nature of the objects of interest. This conjecture needs, of course, to be validated experimentally, which goes beyond the scope of this paper.

\paragraph{Computational complexity}
Although the operations of DST have, in the worst case, exponential complexity, the mass functions computed in the EM module have only $K$ focal sets, where $K$ is the number of classes, and the contextual discounting operation computed in the MMEF is applied to the contour function, as explained in Section \ref{subsubsec:MMEF}. Consequently, the number of operations performed in the EM and MMEF modules is only linear in the number of classes. \revision{More precisely, as shown in \cite{denoeux2000neural}, each forward and backward propagation for one voxel and one modality in the EM module has complexity $O(I(H+K))$, where $I$ is the number of prototypes, $H$ is the number of features extracted by the FE module, and $K$ is the number of classes. In the MMEF module, the discounting  of the $T$ mass functions for each voxel using \eqref{eq:c_discounting_t} and their combination using \eqref{eq:dis_pl} can be performed in $O(KT)$ operations, and the backward pass (gradient calculation) requires the same computational effort. Overall, the complexity of our model is, thus,}  similar to that performed in standard neural network architectures based on weighted sums. In terms of computing times, pre-training each of the FE modules with the nnFormer architecture took approximately one hour on our machine\footnote{\new{All models were trained on an NVIDIA A100-SXM4 graphics card with 40 GB GPU memory.}} for the BraTS2021 dataset, and training the whole system end-to-end took 2.3 hours. The total training time (6.4 hours) is slightly less than that of nnFormer with the four modalities (7.8 hours). As far as state-of-the-art uncertainty quantification techniques are concerned, Monte Carlo dropout does not significantly impact training time, while the deep ensemble method is notoriously time-consuming because it implies training several models. Overall, our framework based on DST and decision fusion is at least as efficient as alternative uncertainty quantification approaches.
 
 \paragraph{Limitations} Our approach is based on combining high-level information extracted from each modality by the FE and EM modules in the form of mass functions. It, thus, has all the advantages and limitations of decision-level fusion approaches. On the plus side, it is highly modular and can still provide sensible results when only some of the modalities are available. This advantage is not crucial in multimodal image segmentation applications because all modalities are usually available, but it can matter in other potential applications such as remote sensing, as mentioned above. Another advantage of decision fusion is that the fusion process is simple and transparent, as already discussed in Sections \ref{subsec: lym}  and \ref{subsec: brain}. On the minus side, decision-level fusion is, at least in principle, suboptimal because it does not consider all input data globally: we can always construct a classification task in which a single classifier trained with a set of features will perform better than a combination of classifiers trained with each of the features. The good performances of our approach reported in Sections \ref{subsec: lym}  and \ref{subsec: brain} show that this potential suboptimality is not an issue in the considered medical image segmentation applications, but it could be in other applications. Another limitation of our approach is that, to keep computations simple, we do not combine the whole discounted mass functions in the MMEF module, but only the contour functions. As a result, the output at each voxel is not a full mass function (with $2^K-1$ focal sets), which prevents us from harnessing the full power of DST, such as some of the decision rules reviewed in \cite{denoeux2019decision}. This and other limitations will be addressed in future work.
 }

\section{Conclusion}
\label{sec:conclu}

We have proposed a deep decision-level fusion architecture for multi-modality medical image segmentation. In this approach, features are first extracted from each modality using a deep neural network such as UNet. An evidence-mapping module based on prototypes in feature space then computes a Dempster-Shafer mass function at each voxel. To account for the varying reliability of different information sources in different contexts, the mass functions are transformed using the contextual discounting operation before being combined by Dempster's rule. The whole framework is trained end-to-end by minimizing a loss function that quantifies prediction error both at the modality level and after fusion.

This model has been evaluated using two real-world datasets for lymphoma segmentation in PET-CT images and brain tumor segmentation in multi-MRI images. In both cases, our approach has been shown to allow for better uncertainty quantification and image segmentation as compared to various alternative schemes based on pixel-level fusion. In particular, as compared to UNet,  nnUNet or nnFormer alone with a softmax layer, the introduction of the evidential mapping module (computing the mass functions) improves the results, and the decision-level fusion scheme with contextual discounting brings an additional improvement. Furthermore, the values found for the reliability coefficients are consistent with domain knowledge, which suggests that these coefficients can provide useful insight into the fusion process.

This work can be extended in many directions. First, as discussed in Section \ref{subsec:discussion}, our DST-based fusion approach can be applied to a variety of learning tasks in which several sources of information must be combined. In the biomedical domain, it could be applied to fuse heterogeneous data such as signals, personal information, biomarkers, gene information, etc. In remote sensing, a potential application could be, e.g., the fusion of Lidar, SAR and hyperspectral data. \new{References \cite{lahat15} and \cite{bokade21} mention many other applications in which multimodal data fusion plays an important role, including human-machine interaction, meteorological monitoring using weather radar and satellite data, or concrete structural monitoring through fusing ultrasonic, impact echo, capacitance, and radar.} From a theoretical point of view, our approach could be extended in several directions. \new{As mentioned in Section \ref{subsec:discussion}, we could combine not only the contour functions from the EM module but the whole mass functions, which would allow us to compute richer outputs that could be exploited within more sophisticated decision strategies \revision{such as partial classification \cite{ma21},} or further combined with other data.} We could also consider other mass-function correction methods making it possible to account for more diverse meta-knowledge about information sources such as proposed, e.g. in \cite{pichon16}, and/or other combination rules such as the cautions rule \cite{denoeux08} or variants with learnable parameters as used in \cite{quost11}.

\section*{Acknowledgements} 
This work was supported by the China Scholarship Council (No. 201808331005). It was carried out in the framework of the Labex MS2T, which was funded by the French Government, through the program ``Investments for the Future'' managed by the National Agency for Research (Reference ANR-11-IDEX-0004-02).

%%%%%%%%%%%%%%%%%%

%%%%%%%%%%%%%%%%%%

%\appendix

%\section{Sample Appendix Section}
%Lorem ipsum dolor sit amet, consectetur adipiscing elit, sed do eiusmod tempor section \ref{sec:sample1} incididunt ut labore et dolore magna aliqua. Ut enim ad minim veniam, quis nostrud exercitation ullamco laboris nisi ut aliquip ex ea commodo consequat. Duis aute irure dolor in reprehenderit in voluptate velit esse cillum dolore eu fugiat nulla pariatur. Excepteur sint occaecat cupidatat non proident, sunt in culpa qui officia deserunt mollit anim id est laborum.

% \bibliographystyle{elsarticle-num} 
 \bibliographystyle{abbrv}
 \bibliography{cas-refs}





% \end{thebibliography}
\end{document}